\documentclass[prd,showpacs,preprintnumbers,amsmath,amssymb]{revtex4}

\usepackage{graphicx}% Include figure files
\usepackage{dcolumn}% Align table columns on decimal point
\usepackage{bm}% bold math

\begin{document}

%\preprint{APS/123-QED}

\title{Tidal Tails Test the Equivalence Principle in the Dark Sector}

\author{Michael Kesden$^1$ and Marc Kamionkowski$^2$}
\affiliation{$^1$Canadian Institute for Theoretical Astrophysics, University of Toronto, Toronto, ON M5S 3H8, Canada\\
$^2$California Institute of Technology, Mail Code 130-33, Pasadena, CA 91125}

\date{\today}

\begin{abstract}
Satellite galaxies currently undergoing tidal disruption offer a unique opportunity to constrain an effective violation of the equivalence principle in the
dark sector.  While dark matter in the standard scenario interacts solely through gravity on large scales, a new long-range force between dark-matter
particles may naturally arise in theories in which the dark matter couples to a light scalar field.  An inverse-square-law force of this kind would
manifest itself as a violation of the equivalence principle in the dynamics of dark matter compared to baryons in the
form of gas or stars.  In a previous paper, we showed that an attractive force would displace stars outwards from the bottom of the satellite's gravitational
potential well, leading to a higher fraction of stars being disrupted from the tidal bulge further from the Galactic center.  Since stars disrupted from the
far (near) side of the satellite go on to form the trailing (leading) tidal stream, an attractive dark-matter force will produce a relative enhancement of
the trailing stream compared to the leading stream.  This distinctive signature of a dark-matter force might be detected through detailed observations of
the tidal tails of a disrupting satellite, such as those recently performed by the Two-Micron All-Sky Survey (2MASS) and Sloan Digital Sky Survey (SDSS) on
the Sagittarius (Sgr) dwarf galaxy.  Here we show that this signature is robust to changes in our models for both the satellite and Milky Way, suggesting
that we might hope to search for a dark-matter force in the tidal features of other recently discovered satellite galaxies in addition to the Sgr dwarf.
\end{abstract}

\maketitle

\section{Introduction} \label{S:intro}
At the dawn of the era of precision cosmology, the nature of the dark sector remains the greatest mystery in cosmology and perhaps all of physics.  By
definition, dark matter and dark energy cannot be observed directly; we must infer their presence and behavior through their influence on the
visible components of the Universe.  Dark matter is ``detected'' through its gravitational contribution to the rotation curves of spiral galaxies and
the velocity dispersion of galaxies in clusters.  Interactions purely within the dark sector that leave dark-matter density profiles largely intact are
particularly difficult to constrain by observations of this kind.  Yet extensions to the standard model of particle physics might incorporate such
dark-matter self-interactions, and an opportunity to constrain them in an astrophysical context could be invaluable.  In a recent {\it Letter} \cite{PRL},
we proposed that the tidal tails of a disrupting satellite galaxy such as the Sagittarius (Sgr) dwarf offer just such an opportunity.  Here we discuss the
numerical techniques used in our simulations of tidal disruption, and show through a new series of simulations that our proposed signature of dark-matter
interactions is robust to changes in our Galactic and Sgr dwarf models.

The signature we are seeking essentially tests the weak equivalence principle, the universality of free-fall between baryons and dark matter in the Galactic
gravitational field.  The luminosity $L_{\rm Sgr} = 1.4 \times 10^7 L_{\odot}$ of the Sgr dwarf attests to its stellar content, while its large velocity
dispersion suggests the presence of considerable amounts of dark matter \cite{2MassTT}.  If the stars and dark matter fall differently, the centers of mass
of the stars and dark matter will separate and stars will no longer be disrupted at roughly equal rates from the tidal bulges on the near and far sides of
the satellite.  Because the typical velocities imparted to stars during tidal disruption are much less than the satellite's orbital velocity about the
Galactic center, these stars will still faithfully trace the satellite's orbit at the time of disruption \cite{TSdyn}.  An attractive dark-matter
interaction would cause the dark matter in the Sgr core to fall faster in the Galactic potential than its stars, leading to a relative enhancement of the
trailing stream formed from stars disrupted on the satellite's far side.  This is a twenty-first century generalization of Galileo's possibly apocryphal test
of the equivalence principle.  Galileo allegedly dropped objects of different mass off the leaning tower of Pisa at the same time and watched to see whether
the heavier object left the lighter behind.  We hope to watch tidally disrupted stars and a dark-matter-dominated satellite race around the Galactic center,
and see whether dark matter with a greater effective value of Newton's constant $G$ will leave the visible matter behind.

Detection of a dark-matter interaction of even a few percent the strength of gravity should be feasible with observations already in hand.  Surface
brightnesses for both the leading and trailing streams of the Sgr dwarf have been observed by both the Two-Micron All-Sky Survey (2MASS) \cite{2MassObs}
and the Sloan Digital Sky Survey (SDSS) \cite{SDSS}.  As stars disrupted from the Sgr dwarf have been identified over the full $360^{\circ}$ of its orbit,
an extremely long lever arm exists over which to make observations.  In this paper, we perform a series of N-body simulations of a satellite with similar
mass and orbit to the Sgr dwarf, and show that a dark-matter interaction 1\% the strength of gravity could reduce the ratio of leading to trailing stars by
33\% or more, while a 4\% dark-matter force could reduce this ratio by over 85\%.  While these results are far from conclusive, the leading-to-trailing
ratio is a crude statistic compared to what could be constructed from the distances and radial velocities of stream stars measured along the Sgr orbit.  The
distances \cite{2MassObs} and radial velocities \cite{2MassVel} to hundreds of M-giant stars in the Sgr streams have already been determined from photometric
parallaxes and spectroscopy.  Furthermore, future missions such as the Space Interferometry Mission (SIM) and Gaia will measure these quantities to
unprecedented precision, and provide proper motions as well.  As this paper is only intended to draw attention to the tight limits that could be set on a
dark-matter force through a more detailed comparison to observations, we restrict our attention here to the leading-to-trailing ratio.

An outline of our theory, methodology, and results is given below.  In Section~\ref{S:DM} we consider how fundamental physics might give rise to a
dark-matter self-interaction, and survey previous attempts to constrain such an interaction in astrophysical settings.  We focus on the case of tidal
streams in Section~\ref{S:tidal}, which have been used in previous research to probe the depth and shape of the host's gravitational potential.  We extend
this work by incorporating a dark-matter self-interaction into pre-existing Galactic models, and propose that its effects are not degenerate with changes to
other more conventional model parameters.  This hypothesis is tested using simulations whose methodology is described in Section~\ref{S:sim}.  The results
of these simulations are presented in Section~\ref{S:res}, with some concluding remarks as to their cosmological implications and hopes for observational
verification given in Section~\ref{S:conc}.

\section{Dark-Matter Interactions} \label{S:DM}

The absence of excess microlensing events towards the Large Magellanic Cloud (LMC) and M31 \cite{MACHO,PG,EROS2} suggests that the majority of dark matter
consists not of compact objects but of a new fundamental particle.  Extensions to the standard model provide possible candidates including sterile
neutrinos, axions \cite{rosenberg,raffelt,turner}, and weakly interacting massive particles (WIMPs) such as supersymmetric neutralinos
\cite{jkg,bergstrom,hooper} and Kaluza-Klein excitations \cite{EPintro}.  All of these dark-matter candidates are expected to have interactions beyond the
purely gravitational, though the nature of these additional interactions is model dependent.  In some theories, new interactions lead to space and time
dependence in the fundamental physical ``constants'' including Newton's constant $G$.  Dynamical measurements of dark matter can in principle be used to
constrain such interactions \cite{EPintro}.  Lacking observations favoring any specific model over another, a dark-matter interaction can be modeled
empirically by a Yukawa coupling between a fermionic dark-matter particle $\psi$ and an additional scalar field $\phi$, described by the Lagrangian
\cite{EPlett},
\begin{equation} \label{E:L1}
L = \Bar{\psi}i\gamma_{\mu}\partial^{\mu}\psi - m_{\psi} \Bar{\psi}\psi + \frac{1}{2} \partial_{\mu} \phi \partial^{\mu} \phi - \frac{1}{2} m_{\phi}^2
\phi^2 + g\phi\Bar{\psi}\psi \, .
\end{equation}
Here and in all equations of this section we are working in natural units where $\hslash = c = 1$.  The Yukawa coupling $g\phi\Bar{\psi}\psi$ of
Eq.~(\ref{E:L1}) is the unique interaction term that is both symmetric under the global transformation $\psi \to \psi e^{i \varphi}$ and has a coupling
constant $g$ of non-negative mass dimension.  These two features lead respectively to conservation of particle number for the dark-matter species $\psi$
and renormalizability \cite{P&S}.  Several of the candidates mentioned above can be fermions, and invoking a scalar field to produce new phenomenology is a
treasured pastime of particle theorists.  Generically, all possible interactions among the constituent fields of a theory should be included in the absence
of a symmetry or other mechanism barring a specific interaction.  This principle implies that new physics including a light scalar might well lead to a
long-range ``fifth force.''

Such a fifth force can theoretically affect visible as well as dark matter.  However, laboratory and Solar System tests of the equivalence principle
prohibit fifth forces between ordinary materials strong enough to be interesting for astrophysics \cite{cliffwill}.  Torsion-balance tests also constrain
long-range interactions between visible and dark matter even when such forces do not operate directly between standard-model particles \cite{stubbs,Adel}.
No such prohibition exists on fifth forces purely in the dark sector, as dark matter has yet to be observed in either the laboratory or Solar System.  As we
still have no compelling reason to prefer a specfic dark-matter interaction over that described in Eq.~(\ref{E:L1}), we will restrict our attention to this
model in the remainder of this paper.  Dark matter in this model consists of fermions $\psi$ of mass $m_{\psi}$, which in the free-field limit $g \to 0$
behave exactly like collisionless WIMPs.  In the nonrelativistic limit, the Yukawa coupling $g\phi\Bar{\psi}\psi$ produces a force between two dark-matter
particles given by the potential \cite{EPlett},
\begin{equation} \label{E:P1}
V_{\phi}(r) = - \frac{g^2}{4\pi r} e^{-m_{\phi} r} \, .
\end{equation}
This potential is the unique solution to the Euler-Lagrange equations following from the Langrangian of Eq.~(\ref{E:L1}) for a point source.  On
scales much less than the Compton wavelength $\lambda \equiv m_{\phi}^{-1}$ of the scalar particle, it yields an attractive, inverse-square-law force just
like gravity, whose potential between two dark-matter particles of mass $m_{\psi}$ is
\begin{equation} \label{E:Pgrav}
V_{\rm grav}(r) = - \frac{G m_{\psi}^2}{r} \, .
\end{equation}
Dividing Eq.~(\ref{E:P1}) by Eq.~(\ref{E:Pgrav}), we see that the Yukawa dark-matter force is suppressed compared to gravity by a factor $\beta^2$, where
\begin{equation} \label{E:betadef}
\beta \equiv \frac{g}{\sqrt{4\pi}}  \left( \frac{m_{\rm Pl}}{m_{\psi}} \right)
\end{equation}
is the dimensionless charge-to-mass ratio, and $m_{\rm Pl} \equiv G^{-1/2} = 1.2 \times 10^{19}$ GeV is the Planck mass.  Note that our $\beta^2$ differs
from $\alpha$ appearing in Ref.~\cite{EPlett} by a factor of $4\pi$ in the denomimator, which was left out of their paper.

This new scalar force could become relevant on galactic or even cosmological scales provided that the scalar mass $m_{\phi}$ is sufficiently small.
The implications of a dark-matter force were first considered in the context of cosmology, where the growth of structure is well described by linear
perturbation theory.  The interaction of Eq.~(\ref{E:L1}) can be incorporated into linear-theory equations in a straightforward manner to yield predictions
for large-scale structure (LSS) including the baryon and dark-matter power spectra \cite{EPlong}.  A repulsive (attractive) DM force would reduce (enhance)
structure formation for $r \lesssim \lambda$, leading to a relative increase (decrease) in large-scale power.  At the time these calculations were first
performed, an $\Omega_m = 1$ universe was strongly favored theoretically and a dark-matter self-interaction was seen as a possible explanation for its
failure to predict adequate large-scale power when normalized to observations at small scales.  Such an interaction also helped to explain observed high rms
bulk velocities, and was consistent with COBE upper bounds on CMB temperature fluctuations.  All observational constraints led to a final bound of
$-0.5 \lesssim \beta^2 \lesssim 1.3$ on the strength of a dark-matter self-interaction for $\lambda \gtrsim$ several hundred kpc \cite{EPlong}.  Negative
values of $\beta^2$ correspond to repulsive dark-matter forces which can be constrained observationally though they are not permitted by the model of
Eq.~(\ref{E:L1}).

Models involving a dark-matter force mediated by a scalar field $\phi$ received renewed attention after the discovery of the cosmic acceleration
\cite{perlmutter,riess}, as it was quickly realized that this acceleration could be produced by a slowly rolling scalar field as in the case of inflation.
This field would generically couple to both visible and dark matter as described above, and thus could only be explained in a unified theory of the dark
sector.  A common explanation for dark matter and dark energy is also desirable because it might solve the cosmic coincidence problem as to why
$\Omega_{\rm DM}$ and $\Omega_{\rm DE}$ are of the same order today despite their different scalings with redshift.  The observed value of a ``plain
vanilla'' cosmological constant also proved embarrassingly difficult to derive on first principles $[ \log (\rho_{\Lambda}/m_{\rm Pl}^{4}) \simeq -120 ]$.
Recent experiments including the Wilkinson Microwave Anisotropy Probe (WMAP) constrained models of coupled dark matter and dark energy, including a
model-dependent limit of $\beta < 0.15$ at 95\% confidence \cite{WMAP}.  This limit is admirably restrictive, but relies on a specific prediction that
$\Omega_{\rm DE}$ track $\Omega_m$ for an extended period \cite{Cquint}.  More generally, some dark-energy models predicted an observationally unacceptable
scale-independent bias $b \equiv \delta_b/\delta_c$ less than 0.73 in regions of parameter space that solved the coincidence problem \cite{bias}.
Increasingly sophisticated theories of coupled dark energy with several dark-matter species were developed to avoid this constraint \cite{interact}.  In one
theory, a second, massless dark-matter species drives the scalar field $\phi$ to zero except in the most massive halos, screening any $\phi$-mediated
dark-matter forces on larger scales.  After the second species is sufficiently diluted by the cosmic expansion, $\phi$ can vary and act like the dark energy
for appropriate choices of its potential $V(\phi)$.  A second theory motivated by string theory replaces the WIMP $\psi$ by two oppositely charged species
$\psi_{\pm}$ to preserve global scalar-charge neutrality \cite{string}.  Large-scale modes are purely adiabatic (equal contributions from both species) by
design, and neither source nor respond to a scalar field in linear perturbation theory.  However the two species are subject to charge separation on
nonlinear scales, which is expected to be complete in well-relaxed halos for gravitational-strength dark-matter forces \cite{STcosmo}.  While there are
numerous other variations on this theme, the main point to take away is that there are many models in which the dark-matter force is insignificant on
cosmological scales but potentially important on scales at which the growth of structure has gone nonlinear.

One application of a dark-matter force on nonlinear scales that served as an initial motivation for this paper was as a possible solution to a lack of
dwarf galaxies within voids \cite{voids}.  Simulations of structure formation in $\Lambda$CDM universes indicate the presence of low-mass halos within
voids in the distribution of larger galaxies that should be massive enough to host dwarf galaxies in these regions.  Some have claimed that too few dwarf
galaxies have actually been observed within voids, and that this deficit is too large to be explained by conventional means such as feedback from
supernovae or an ionizing background.  An attractive dark-matter self-interaction will more effectively clean out voids in the distribution of larger
galaxies, as well as limit accretion from the IGM in better agreement with observations.  Simulations of structure in a 50 $h^{-1}$ box were performed using
the dark-matter potential of Eq.~(\ref{E:P1}) indicating that values of $\beta \gtrsim 1$ and $m_{\phi}^{-1} \gtrsim 1 h^{-1}$ Mpc were neccessary to
adequately address the void problem \cite{nusser}.

While the cutoff for $r \gtrsim m_{\phi}^{-1}$ might explain how a dark-matter force strong enough to solve the void problem could be consistent with
large-scale surveys like SDSS \cite{Verde}, we wondered whether it could be reconciled with observations on the scale of individual halos.  In the presence
of an inverse-square-law dark-matter force, dark-matter particles effectively gravitate differently from stars or gas, as if one had artificially tuned $G$
to a different value in violation of the equivalence principle.  Stars used to trace a galaxy's rotation curve will interact purely gravitationally with
the dark matter, leaving galactic estimates of dark-matter density profiles unchanged.  On cluster scales however, where individual galaxies with their own
dark-matter halos are used as tracers of the cluster density profile, the additional force due to an attractive dark-matter self-interaction would be
falsely attributed to additional gravitating cluster mass \cite{EPlett}.  When combined with unbiased estimates of the cluster mass, derived from weak
lensing or X-ray gas temperatures, this enhancement of the velocity dispersion of cluster galaxies can in principle be used to constrain a dark-matter
force.  In practice, it is doubtful whether systematic errors associated with either of these methods will allow such a comparison.  A second cluster test
for a dark-matter force involves the baryon-to-dark-matter ratio, as baryons are preferentially lost compared to more tightly bound dark-matter particles
during the mergers in which the cluster is formed \cite{clusters}.  A dark-matter force can decrease the cluster baryon-to-dark-matter ratio by 10\% for
$\beta$ as low as 0.06.  However, even in the absence of a dark-matter force, the baryon-to-dark-matter ratio in clusters is expected to be lower than the
cosmological $\Omega_b/\Omega_c$ because of the still uncertain physics of cluster gas.

The impressive limits set by cluster tests suggest that small-scale structure may indeed be the most promising venue in which to search for an
equivalence-principle-violating force.  With as speculative a conjecture as a dark-matter self-interaction, one would ideally like a true ``smoking gun'', a
signature that could not be mimicked by more conventional changes to cluster masses or density profiles.  In Ref.~\cite{PRL}, we proposed that an asymmetry
in the stellar densities along the leading and trailing tidal streams of a dark-matter-dominated satellite galaxy constituted just such a smoking gun.  The
simulations we presented there showed that $\beta$ as low as 0.2 can lead to a pronounced suppression of the leading stream.  In this paper, we examine just
how robust this signature is to changes in our Milky Way model and the mass and orbit of the satellite galaxy.  This study should help us to determine how
well other recently discovered satellite galaxies besides the Sgr dwarf can be used to constrain dark-matter forces.  To do this, we first take a closer
look at the theory of tidal disruption in the next section.

\section{Tidal Disruption} \label{S:tidal}

Astronomers have long been fascinated by the spectacular bridges and tails observed to connect certain closely separated pairs of galaxies, NGC 4038/39
(the Antennae) and NGC 4676 (the Mice) being among the most famous examples.  In a classic paper, Toomre and Toomre showed in a series of
simulations that a parabolic encounter between a satellite galaxy and a more massive partner can generate both a long and curving tidal tail and extensive
near-side debris \cite{Toomre}.  Disk particles in these simulations did not self-gravitate, but merely traced orbits in the potentials of the two galaxies
which were modeled as point particles.  The success with which these simulations can reproduce many of the features observed in real systems suggested that
these features were more sensitive to the orbits in the host galaxy's potential than either the precise details of the disruption event itself or the
internal structure of the satellite.   The physics behind this result is illuminated by the energetics of the host-satellite system, which is characterized
by the orbital energy $E_{\rm orb}$, the energy $E_{\rm tid}$ imparted during tidal disruption, and the internal binding energy $E_{\rm bin}$ of the
satellite \cite{TSdyn}.  For a satellite of mass $m_{\rm sat}$ and radius $r_{\rm sat}$ on an orbit of semi-major axis $R$ enclosing a mass $M_R$ of its
host, the tidal radius $r_{\rm tid}$ can be approximated as
\begin{equation} \label{E:rtid}
r_{\rm tid} = R \left( \frac{m_{\rm sat}}{M_R} \right)^{1/3} \, .
\end{equation}
Tidal disruption begins when the satellite fills its Roche lobe, $r_{\rm sat} = r_{\rm tid}$, at which point the three energy scales are given by
\begin{eqnarray} \label{E:3E}
E_{\rm orb} &=& \frac{GM_R}{R} \\
E_{\rm tid} &=& r_{\rm tid} \frac{d\Phi_{\rm host}}{dR} = \left( \frac{m_{\rm sat}}{M_R} \right)^{1/3} E_{\rm orb} \\
E_{\rm bin} &=& \frac{Gm_{\rm sat}}{r_{\rm sat}} = \left( \frac{m_{\rm sat}}{M_R} \right)^{2/3} E_{\rm orb} \, .
\end{eqnarray}
For Galactic tidal streams, the satellite is much less massive than its host, $m_{\rm sat}/M_R \ll 1$, implying a well-defined hierarchy in the energy
scales
\begin{equation} \label{E:Ehier}
E_{\rm orb} \gg E_{\rm tid} \gg E_{\rm bin} \, .
\end{equation}
The first inequality shows why the tidal debris remains on orbits similar to that of the satellite itself, while the second indicates that the satellite
should have little influence on the debris after it has become unbound.

The persistence of tidal debris in distinctive streams over several orbital periods can be understood by realizing that the debris disperses on the
``dephasing'' time $t_{\rm deph}$ of adjacent orbits in the host potential, which in general will be much longer than the typical dispersal time
$t_{\rm disp}$ given by the extent of the debris $r_{\rm tid}$ divided by its internal velocity dispersion $v_{\rm tid} \equiv E_{\rm tid}^{1/2}$
\cite{Kuhn}.  For a host potential with orbital angular velocities $\omega (R)$ and dynamical time
\begin{equation} \label{E:tdyn}
t_{\rm dyn} = \left( \frac{GM_R}{R^3} \right)^{-1/2} \, ,
\end{equation}
the two timescales for dispersal of tidal debris are given by
\begin{eqnarray} \label{E:2t}
t_{\rm deph} &=& \left( \left| \frac{d\omega}{dR} \right| r_{\rm tid} \right)^{-1} = \left( \frac{m_{\rm sat}}{M_R} \right)^{-1/3} t_{\rm dyn} \\
t_{\rm disp} &=& \frac{r_{\rm tid}}{v_{\rm tid}} = \left( \frac{m_{\rm sat}}{M_R} \right)^{1/6} t_{\rm dyn} \, .
\end{eqnarray}
We see that $t_{\rm deph} \gg t_{\rm disp}$ for small satellite-to-host mass ratios, and that the dephasing times will be even greater for NFW halos whose
rotation curves are much flatter than those of a Kepler potential.  Assuming best-fit values of $t_{\rm dyn} = 0.85$ Gyr and
$m_{\rm sat}/M_R \simeq 10^{-3}$ for the Sgr dwarf \cite{2MassTT}, we find $t_{\rm deph} \simeq 8.5$ Gyr confirming that the Sgr dwarf can survive for many
orbits about the Galactic center to develop extensive tidal streams.

The energetics argument presented above can help explain the morphology of tidal streams in addition to the timescale for their formation.  As the satellite
fills its Roche lobe, stars are disrupted from the sides closest and most distant from the Galactic center.  This leads to a bimodal distribution in the
energies of the disrupted stars peaked at $E_{\rm orb} \pm E_{\rm tid}$.  Stars disrupted from the near side lose energy and fall deeper in the host's
potential, while those disrupted from the far side gain energy and consequently are boosted onto higher orbits.  Since for all reasonable Galactic
potentials $\omega (R)$ is a decreasing function of $R$, the stars that lose energy race ahead of the satellite's still bound core eventually forming a
leading tidal stream.  Conversely, stars disrupted from the far side of the satellite move to orbits on which they lag behind the satellite and develop into
a trailing tidal tail.

This scenario for the formation of tidal streams is significantly altered by an equivalence-principle-violating force exerted on the satellite's dark-matter
halo by that of the host galaxy.  As the satellite's halo experiences a supplemental acceleration due to the dark-matter force, its center-of-mass is
displaced with respect to that of the satellite's bound stars.  For a satellite on a circular orbit, the additional centripetal acceleration will cause the
halo to orbit the Galaxy more quickly than it would at the same radius $R$ in the absence of a dark-matter force.  The speed will be increased by a factor
\begin{equation} \label{E:fv}
f_v \equiv \sqrt{1 + \beta^2 f_{\rm sat} f_R} \, ,
\end{equation}
where
$f_{\rm sat}$ and $f_R$ are, respectively, the dark-matter fractions of the satellite and the host interior to $R$.  The stars will be displaced to larger
$R$ so that the gravitational pull of the additional interior satellite mass can provide the extra centripetal force necessary to carry the stars around the
Galactic center at this higher speed.  For satellites on eccentric orbits like the Sgr dwarf, the stars will be similarly displaced {\it behind} the
center of mass of the dark-matter halo as it accelerates inwards towards the Galactic center,  and {\it ahead} of the dark-matter halo as it decelerates
during the outward portion of the orbit.  This is entirely analogous to passengers in a car being pressed back against the seats as the car accelerates and
forward against the seat belts as the car decelerates.  The key point to note is that in all instances the stars are displaced to larger $R$ and thus higher
in the Galactic potential well.  From this displaced position, stars will be preferentially disrupted from the side of the satellite furthest from the
Galactic center.  The bimodal distribution in the energies of disrupted stars will now be more strongly peaked about $-E_{\rm orb} + E_{\rm tid}$ than
$-E_{\rm orb} - E_{\rm tid}$, resulting in higher stellar surface densities in the trailing as compared to the leading tidal stream.

The magnitude of this effect can be seen in an analytical estimate of the tidal radii.  For simplicity, we consider a point mass $m_{\rm sat}$ on a
purely radial orbit about a larger point mass $M_R$.  Even in the absence of a dark-matter force, there will be an asymmetry between the tidal radii on the
near and far sides of the satellite because of the non-uniformity in the gradient of the Galactic potential.  The tidal radius is defined to be the distance
from the satellite at which the relative acceleration between a star and the satellite vanishes.  On the near and far sides of the satellite,
\begin{eqnarray} \label{E:RAnear}
\Delta a_{\rm near} &=& a_{\ast} - a_{\rm sat} = -\frac{GM_R}{(x - r_{\rm tid})^2} + \frac{Gm_{\rm sat}}{r_{\rm tid}^2} + \frac{GM_R}{x^2} = 0 \\
\Delta a_{\rm far} &=& a_{\ast} - a_{\rm sat} = -\frac{GM_R}{(x + r^{\prime}_{\rm tid})^2} - \frac{Gm_{\rm sat}}{r^{\prime 2}_{\rm tid}} +
\frac{GM_R}{x^2} = 0 \, , \label{E:RAfar}
\end{eqnarray}
where $x$ is the distance between the satellite and host galaxies and $r_{\rm tid}$ $(r^{\prime}_{\rm tid})$ is the tidal radius on the near (far) side of
the satellite.  Defining $u \equiv r_{\rm tid}/x$ and $u' \equiv r^{\prime}_{\rm tid}/x$, we can rearrange Eqs.~(\ref{E:RAnear}) and (\ref{E:RAfar}) to
obtain
\begin{eqnarray} \label{E:unear}
u^3 &=& \frac{m_{\rm sat}}{M_R} \frac{1 - 2u + u^2}{2 - u} \\
u'^3 &=& \frac{m_{\rm sat}}{M_R} \frac{1 + 2u' + u'^2}{2 + u'} \, . \label{E:ufar}
\end{eqnarray}
In the limit $m_{\rm sat} \ll M_R$, $u, u' \ll 1$ and we can set them to zero on the right-hand sides of Eqs.~(\ref{E:unear}) and (\ref{E:ufar}) to find the
zeroth-order result
\begin{equation} \label{E:zeroth}
u = u' = \left( \frac{m_{\rm sat}}{2M_R} \right)^{1/3} \equiv r_0 \, .
\end{equation}
Expanding the right-hand sides of Eqs.~(\ref{E:unear}) and (\ref{E:ufar}) to first order in $u, u'$, we find
\begin{eqnarray} \label{E:FOnear}
u^3 &\simeq& \frac{m_{\rm sat}}{2M_R} \left( 1 - \frac{3u}{2} \right) \\
u'^3 &\simeq& \frac{m_{\rm sat}}{2M_R} \left( 1 + \frac{3u'}{2} \right) \, . \label{E:FOfar}
\end{eqnarray}
Inserting the zeroth-order result $u = u' = r_0$ into the right-hand sides of Eqs.~(\ref{E:FOnear}) and (\ref{E:FOfar}) leads to first-order corrections to
the near and far side tidal radii
\begin{eqnarray} \label{E:NATnear}
u &\simeq& r_0 - r_1 \\
u' &\simeq& r_0 + r_1 \, , \label{E:NATfar}
\end{eqnarray}
where $r_1 = 0.5(m_{\rm sat}/2M_R)^{2/3}$.  This leads a natural asymmetry of 
\begin{equation} \label{E:DelNAT}
\Delta r_{\rm nat} \equiv u' - u = 2r_1 = \left( \frac{m_{\rm sat}}{2M_R} \right)^{2/3}
\end{equation}
between the near and far side tidal radii which serves as a proxy for the asymmetry in the stellar densities along the leading and trailing tidal streams.

We wish to compare this asymmetry $\Delta r_{\rm nat}$ with the asymmetry $\Delta r_{\rm DM}$ induced by a dark-matter force to determine when the latter
will be dominant.  This dark-matter asymmetry can be calculated by adding the acceleration produced by the dark-matter force to the satellite's acceleration
appearing in Eqs.~(\ref{E:RAnear}) and (\ref{E:RAfar}).  This leads to 
\begin{eqnarray} \label{E:DMAnear}
\Delta a_{\rm near} &=& a_{\ast} - a_{\rm sat} = -\frac{GM_R}{(x - r_{\rm tid})^2} + \frac{Gm_{\rm sat}}{r_{\rm tid}^2} + \frac{GM_R}{x^2}
(1 + \beta^2 f_{\rm sat} f_R) = 0 \\
\Delta a_{\rm far} &=& a_{\ast} - a_{\rm sat} = -\frac{GM_R}{(x + r^{\prime}_{\rm tid})^2} - \frac{Gm_{\rm sat}}{r^{\prime 2}_{\rm tid}} +
\frac{GM_R}{x^2} (1 + \beta^2 f_{\rm sat} f_R) = 0 \, . \label{E:DMAfar}
\end{eqnarray}
Defining $u, u'$ as previously and assuming they are much less than unity, we find
\begin{eqnarray} \label{E:FOBnear}
u^3 &\simeq& \frac{m_{\rm sat}}{2M_R} \left( 1 + \frac{\beta^2 f_{\rm sat} f_R M_R}{m_{\rm sat}} u^2 \right) \\
u'^3 &\simeq& \frac{m_{\rm sat}}{2M_R} \left( 1 - \frac{\beta^2 f_{\rm sat} f_R M_R}{m_{\rm sat}} u'^2 \right) \, , \label{E:FOBfar}
\end{eqnarray}
leading to corrections to $u, u'$ of the form of Eqs.~(\ref{E:NATnear}) and (\ref{E:NATfar}).  To first order in $\beta^2$, these corrections yield a
dark-matter asymmetry of
\begin{equation} \label{E:DelDM}
\Delta r_{\rm DM} \equiv u' - u = 2r_1 = -\frac{1}{3} \beta^2 f_{\rm sat} f_R \, .
\end{equation}
The dark-matter asymmetry $\Delta r_{\rm DM}$ is negative because an attractive dark-matter force fosters tidal disruption from the far side of the
satellite, reducing $u'$ and enhancing $u$.  By contrast, the natural asymmetry $\Delta r_{\rm nat}$ is positive because the Galactic potential is steeper
on the near side of the satellite.  Equating the two asymmetries, we see that the dark-matter asymmetry will dominate when
\begin{equation}
\beta^2 > \frac{3}{f_{\rm sat} f_R} \left( \frac{m_{\rm sat}}{2M_R} \right)^{2/3} \, ,
\end{equation}
which for the default model parameters we use for the Sgr dwarf corresponds to $\beta = 0.21$.  Our simulations reveal a slightly greater sensitivity to a
dark-matter force, probably because of the flatter Galactic potential and eccentric orbit, but this argument successfully captures the magnitude of the
effect we are considering.

In addition to the asymmetry produced during tidal disruption as discussed above, the dark-matter force creates further asymmetry {\it after} the stars
have been tidally disrupted from the satellite.  Some stars will manage to lose energy despite an attractive dark-matter force, but will no longer\
necessarily go on to form a leading stream as previously.  If the amount of energy lost is less than
\begin{equation} \label{E:Emin}
\Delta E_{\rm max} = \frac{1}{2} \beta^2 f_{\rm sat} f_R E_{\rm orb} \, ,
\end{equation}
the disrupted stars will still have a larger kinetic energy than those on a purely gravitational circular orbit at radius $R$, and will consequently move
to higher, longer-period orbits in the Galactic potential well.  Even stars that lose the full $\Delta E_{\rm max}$ during tidal disruption will move on
their circular orbits with velocities lower than that of the dark-matter-dominated satellite by the factor $f_v$ of Eq.~(\ref{E:fv}).  This reduction in
speed can only cause the disrupted stars to fall behind by an additional $2\pi f_v \simeq \pi \beta^2 f_{\rm sat} f_R$ radians per orbit, an amount
insufficient to explain the magnitude of the effect seen in Ref.~\cite{PRL} and Section~\ref{S:res} of this paper.  We see that the dominant effect is the
displacement of the satellite's stars with respect to their dark-matter halo and the subsequent asymmetry between disruption from the satellite's near and
far sides.  It is the narrowness in the two peaks in the energy distribution of tidally disrupted stars, following from the second inequality
$E_{\rm tid} \gg E_{\rm bin}$ of Eq.~(\ref{E:Ehier}), that leads to the surprising sensitivity of our proposed test for a dark-matter force.

The theory developed above reveals that even in the absence of a dark-matter force, we can learn about the depth and shape of the Galactic potential through
the dependence of tidal-stream morphology on $M_R$ and $\omega(R)$.  This idea was first put into practice by Lynden-Bell, who used the assumption that the
Fornax, Leo I, Leo II, and Sculptor dwarf spheroidal (dSph) galaxies were all remnants of a larger, tidally disrupted satellite to estimate that the Milky
Way had a mass of $4.6 \times 10^{11} M_{\odot}$ within 85 kpc of its center \cite{LB}.  Later work sought to apply the constraint
$t_{\rm deph} > t_{\rm dyn}$ to the Ursa Minor and Draco dSph galaxies to obtain similar estimates for the mass of the Galaxy and the shape of its potential
\cite{Kuhn}.  The discovery of the Sgr dwarf galaxy \cite{disc} and its extensive stellar stream \cite{stream} at a distance of only 24 kpc provided an
opportunity to test theories of tidal disruption with unprecedented precision.  While initial work based solely on the bound core of the satellite failed to
place tight constraints on the Galactic halo \cite{JSH95}, later studies making use of carbon stars in the stream with measured distances and radial
velocities identified the plane of the Sgr orbit and set limits on the oblateness of the Galactic halo \cite{CS1,CS2}.  However even these observations were
still consistent with both purely stellar models of the Sgr dwarf $(M/L = 2.25)$ and those with an extended dark-matter halo \cite{HelWhi01}.  This is
partly because as we have seen the tidal streams are far more sensitive to orbits in the Galactic halo than the internal dynamics of the satellite itself.
The far more numerous M-giant stars discovered by 2MASS were able to provide the most accurate determination yet of parameters of the Sgr and its orbit
\cite{2MassObs}.  A comparison of these observations with simulations found $M_{\rm Sgr} = (2-5) \times 10^8 M_{\odot}$, $M_{\rm Sgr}/L_{\rm Sgr} = 14$ to
36, and a Sgr orbit with pericenter 10 to 19 kpc, apocenter 56 to 59 kpc, and period 0.85 to 0.87 Gyr \cite{2MassTT}.  We use these parameters as
inspiration for the simulations described in the next section, though we do not attempt to reproduce detailed features in the observed stellar streams.

\section{Simulations} \label{S:sim}

The extraordinary development of computer technology has made N-body simulations an invaluable tool for astrophysics, particularly for highly nonlinear,
nonsymmetric problems such as the tidal disruptions considered here.  The numerical algorithms used to implement these simulations have improved
dramatically as well, in part to more efficiently exploit these greatly expanded computing resources.  All of the simulations described in this paper
were performed using a modified version of GADGET-2 (GAlaxies with Dark matter and Gas intEracT), a publicly available cosmological simulation code
developed by Volker Springel \cite{Gadget2}.  Those interested in the detailed workings of GADGET-2 are strongly encouraged to consult his paper
\cite{Gadget2} and the online user guide, but we will briefly summarize the features of the code we used and the changes that were made to incorporate
dark-matter forces.  As its name implies, GADGET-2 can be used to simulate gas in addition to collisionless stars and dark-matter particles.  In principle,
we could use this capability to simulate the {\it gaseous} tidal tail of the LMC which should be sensitive to dark-matter forces like the {\it stellar}
streams of the Sgr dwarf.  However, ram pressure from a low-density ionized Galactic halo could preferentially sweep gas behind the LMC core \cite{LMCgas}
mimicking the signature we are seeking for an attractive dark-matter force.  For this reason, in addition to uncertainties in the feedback affecting gas in
astrophysical settings and continuing discrepancies between numerical methods, we restrict our attention to collisionless stars and dark-matter particles
which can be simulated exclusively with GADGET-2's routines for gravitational forces.

The primary method employed by GADGET-2 to calculate gravitational forces is a recursive tree algorithm in which particles are hierarchically grouped into
larger and larger cells \cite{Gadget2}.  In this approach, pioneered by Barnes and Hut \cite{BarHut}, the cubic box in which the simulation takes place is
divided in half in each direction, resulting in eight daughter nodes.  Each of the particles is assigned to a daughter node, then the daughter nodes are
further subdivided until a hierarchical ``oct-tree'' is created whose lowest-level nodes each contain a single particle.  The center of mass for each node
in the oct-tree is calculated and stored during tree construction.  When calculating the force on a given particle, a node of mass $M$, length $l$, and
distance $r$ is only opened if
\begin{equation} \label{E:OC}
\frac{GM}{r^2} \left( \frac{l}{r} \right)^2 \geq \alpha |{\textbf a}| \, ,
\end{equation}
where $\alpha$ is a small, dimensionless parameter and ${\textbf a}$ is the acceleration of the previous timestep.  Otherwise, the force from that node is
approximated by its monopole moment, that of a point particle of mass $M$ located at the center of mass.  This reduces the number of force calculations per
particle from $N-1$ to $\mathcal{O}(\log N)$.  The opening criterion of Eq.~(\ref{E:OC}) offers advantages over the simple geometrical condition in which
all nodes are opened that subtend an angle $l/r$ greater than some particular value $\theta$.  In the case of a highly symmetric distribution, nodes to the
left and right exert large forces of almost equal magnitude and opposite direction, and the criterion of Eq.~(\ref{E:OC}) ensures that enough nodes are
opened for these partial forces to cancel properly.  Although the oct-tree is not reconstructed from scratch at each timestep $\Delta t$, the monopole
moment of each node is adjusted by drifting the position of the center of mass by an amount $v \Delta t$.

The scalar potential of Eq.~(\ref{E:P1}) leads to an inverse-square-law force on scales $r \ll m_{\phi}^{-1}$ allowing us to use this same recursive tree
algorithm to calculate our dark-matter forces.  GADGET-2 can accommodate up to six different particle species, each with its own mass and gravitational
softening length.  We adapted the program to include a charge-to-mass ratio $\beta$ for each species as well.  Unlike electromagnetism, scalar forces are
attractive between charges of the same sign and repulsive for those of opposite sign.  As the model of Eq.~(\ref{E:L1}) includes only a single dark-matter
species, we will mostly consider attractive dark-matter forces, though our code can just as easily simulate particles with charges of opposite sign.  A
simulation of a tidal disruption with a repulsive dark-matter force between Galactic and satellite halo particles of opposite sign is presented in
Subsection~\ref{SS:rep}.  Each node of the gravitational oct-tree will now include not just a center-of-mass position and velocity, but a center-of-charge
position and velocity for each charged species.  Nodes are opened during the force calculation whenever the criterion of Eq.~(\ref{E:OC}) is exceeded for
gravity or the dark-matter force generated by {\it any} of the charged species.  This acceleration criterion is even more important for dark-matter forces
than for gravity, as initial adiabatic perturbations composed of equal numbers of oppositely charged particles may produce large partial forces that
precisely cancel.  Tracking the center of charge of each species also allows us accomodate the charge separation expected to occur on nonlinear scales.

Appropriately modifying GADGET-2 to include dark-matter forces is vital to our investigation, but equally important is a suitable choice of models for the
host and satellite galaxies.  Our simulations will not yield reliable results unless the initial conditions capture the essential features of the system
under consideration.  A realistic Galactic model will include distinct distributions for bulge, disk, and halo particles.  These distributions must be close
enough to a self-consistent equilibrium solution to the coupled Poisson and collisionless Boltzmann equations to be stable for the duration of the
simulation.  Early attempts at N-body realizations of compound galaxies chose simple, observationally motivated forms for the density profiles of the bulge,
disk, and halo particles.  In what is known as the local Maxwellian approximation, the velocity distributions for these profiles are then approximated as
multivariate Gaussians, with dispersions given by moments of the collisionless-Boltzmann equations and cutoffs set to the local escape velocity
\cite{HernComp}.  This approach is acceptable in many situations, but as the initial configurations are not genuine solutions to the collisionless-Boltzmann
equations they will include transient perturbations and may relax to solutions that differ significantly from the real systems one intended to simulate.  In
particular, the cusps of NFW profiles develop cores on less than a dynamical time and initially isotropic velocity distributions become biased towards radial
orbits \cite{Maxbad}.  These effects cause satellite galaxies to undergo more rapid tidal disruption, as their particles are less tightly bound and spend
more time near the apocenters of their orbits.  As we do not want such unphysical behavior to interfere with our simulations of tidal disruption, we must
adopt a more sophisticated approach that goes beyond the local Maxwellian approximation.

This alternative approach begins with separate distribution functions for the bulge, disk, and halo that are analytic functions of the
the energy $E$ and angular momentum $L_z$ in the $z$-direction \cite{KuDub}.  Both $E$ and $L_z$ are integrals of motion for static, axisymmetric systems.
Jean's theorem guarantees that such distribution functions are solutions of the collisionless-Boltzmann equations and yield potential-density pairs given by
the Poisson equation \cite{B&T}
\begin{equation} \label{E:Pois}
\nabla^2 \Phi = 4\pi G \rho = 4\pi G \int \, f d^3 {\textbf v} \, .
\end{equation}
The bulge in our models is described by the standard Hernquist profile \cite{HernBul},
\begin{equation} \label{E:bulge}
\rho_{\rm H}(r) = \frac{\rho_b}{(r/a_b)(1 + r/a_b)^3} \, ,
\end{equation}
while the halo closely approximates an Navarro-Frenk-White (NFW) profile \cite{NFW},
\begin{equation} \label{E:NFW}
\rho_{\rm NFW}(r) = \frac{\rho_h}{(r/a_h)(1 + r/a_h)^2} \, .
\end{equation}
These profiles extend to arbitrarily large $r$ making them formally unsuitable for N-body simulations, but this can be remedied by imposing cutoffs in the
distribution functions above energy thresholds $E_h, E_b < 0$ \cite{WidDub}.  The disk distribution function depends on $E$, $L_z$, and an approximate third
integral $E_z$ and produces a density profile that behaves like an exponential disk \cite{KuDub}
\begin{equation} \label{E:disk}
\rho_{\rm disk}(R,z) = \frac{M_d}{4\pi R_{d}^2 z_d} e^{-R/R_d} {\rm sech}^2 \left( \frac{z}{z_d} \right) \, ,
\end{equation}
for small $R$, but cuts off at a radius $R_{\rm out}$ on a scale $\delta R_{\rm out}$.

Combining the bulge, disk, and halo of Eqs.~(\ref{E:bulge}) through (\ref{E:disk}) into a self-consistent composite model is nontrivial, as the distribution
functions entering on the right-hand side of Eq.~(\ref{E:Pois}) depend on $E$ and thus $\Phi$ which is sourced by all three components.  Simply replacing
the energy in isolation $E'$ with the energy $E$ in the combined potential leads to composite models in which the components look very different together
than they did separately \cite{WidDub}.  By adopting analytic mappings between $E$ and $E'$ with the appropriate asymptotic behavior, composite models close
to the desired result can be obtained that are still solutions to the collisionless-Boltzmann equations.  This is the approach employed by the program
GALACTICS developed by John Dubinski and Larry Widrow that we used to generate our initial N-body distributions \cite{WidDub}.  A nonzero charge-to-mass
ratio $\beta$ was incorporated into these initial conditions by augmenting the initial velocities of dark-matter particles at a distance $r$ from the
Galactic center by a factor $\sqrt{1 + f_{\rm DM}(r) \beta^2}$, where $f_{\rm DM}(r)$ is the mass fraction of dark matter within $r$.  Although this
approximation is crude, it preserved the initial density profile of the dark matter which appeared stable over the course of the simulation.

\section{Results} \label{S:res}

\begin{table}[t]
\begin{center}
\begin{tabular}{ | c | c | c | c | c | c | c | c | c | c | c |}
\hline
Run & $\beta$ & $\tau$ (in Gyr) & $a_h$ (in kpc) & $r_{\rm tid}$ (in kpc) & $M_{\rm Sgr}$ (in $ 10^9 M_{\odot}$) & $S_{\rm Sgr}$ & Orbit & MW model & Sgr model & $M/L$ (in $M_{\odot}/L_{\odot}$) \\ \hline
1a  & 0.0     & 1.06            & 1.14           & 4.91                   & 0.5                                  & 0             & polar & b        & MTL       & 40 \\ \hline
1b  & 0.1     & 1.05            & 1.13           & 4.97                   & 0.5                                  & 0             & polar & b        & MTL       & 40 \\ \hline
1c  & 0.2     & 1.04            & 1.13           & 5.15                   & 0.5                                  & 0             & polar & b        & MTL       & 40 \\ \hline
1d  & 0.3     & 1.02            & 1.11           & 5.45                   & 0.5                                  & 0             & polar & b        & MTL       & 40 \\ \hline
2a  & 0.0     & 1.06            & 1.44           & 6.16                   & 1.0                                  & 0             & polar & b        & MTL    	& 40 \\ \hline
2b  & 0.1     & 1.05            & 1.43           & 6.21                   & 1.0                                  & 0             & polar & b        & MTL	& 40 \\ \hline
2c  & 0.2     & 1.04            & 1.43           & 6.37                   & 1.0                                  & 0             & polar & b        & MTL	& 40 \\ \hline
3a  & 0.0     & 1.06            & 1.14           & 4.91                   & 0.5                                  & +1            & polar & b        & MTL	& 40 \\ \hline
3b  & 0.2     & 1.04            & 1.13           & 5.15                   & 0.5                                  & +1            & polar & b        & MTL	& 40 \\ \hline
3c  & 0.0     & 1.06            & 1.14           & 4.91                   & 0.5                                  & $-1$          & polar & b        & MTL	& 40 \\ \hline
3d  & 0.2     & 1.04            & 1.13           & 5.15                   & 0.5                                  & $-1$          & polar & b        & MTL	& 40 \\ \hline
4a  & 0.0     & 1.06		& 1.14		 & 4.91			  & 0.5					 & 0		 & circular & b	    & MTL	& 40 \\ \hline
4b  & 0.2     & 1.04		& 1.13		 & 5.15			  & 0.5					 & 0		 & circular & b     & MTL	& 40 \\ \hline
5a  & 0.0     & 1.06            & 1.14           & 4.91                   & 0.5                                  & 0             & planar & b       & MTL	& 40 \\ \hline
5b  & 0.1     & 1.05            & 1.13           & 4.97                   & 0.5                                  & 0             & planar & b       & MTL	& 40 \\ \hline
5c  & 0.2     & 1.04            & 1.13           & 5.15                   & 0.5                                  & 0             & planar & b       & MTL	& 40 \\ \hline
6a  & 0.0     & 1.29            & 1.01           & 5.57                   & 0.5                                  & 0             & polar & a        & MTL	& 40 \\ \hline
6b  & 0.1     & 1.29            & 1.01           & 5.62                   & 0.5                                  & 0             & polar & a        & MTL	& 40 \\ \hline
6c  & 0.2     & 1.27            & 1.00           & 5.76                   & 0.5                                  & 0             & polar & a        & MTL	& 40 \\ \hline
7a  & 0.0     & 1.06            & 1.14           & 4.91                   & 0.5                                  & 0             & polar & b        & 100\% MB  & 40 \\ \hline
7b  & 0.2     & 1.04            & 1.13           & 5.15                   & 0.5                                  & 0             & polar & b        & 100\% MB  & 40 \\ \hline
7c  & 0.0     & 1.06            & 1.14           & 4.91                   & 0.5                                  & 0             & polar & b        & 25\% MB   & 40 \\ \hline
7d  & 0.2     & 1.04            & 1.13           & 5.15                   & 0.5                                  & 0             & polar & b        & 25\% MB   & 40 \\ \hline
8a  & 0.0     & 1.06            & 1.14           & 4.91                   & 0.5                                  & 0             & polar & b        & MTL	& 10 \\ \hline
8b  & 0.2     & 1.04            & 1.13           & 5.10                   & 0.5                                  & 0             & polar & b        & MTL	& 10 \\ \hline
8c  & 0.0     & 1.06            & 1.14           & 4.91                   & 0.5                                  & 0             & polar & b        & MTL	& 4.5 \\ \hline
8d  & 0.2     & 1.05            & 1.13           & 5.04                   & 0.5                                  & 0             & polar & b        & MTL	& 4.5 \\ \hline
9   & -0.2    & 1.04            & 1.13           & 5.15                   & 0.5                                  & 0             & polar & b        & MTL	& 40 \\ \hline
\end{tabular}
\caption{Model parameters for each simulation.  From left to right, they are: the dark-matter charge-to-mass ratio $\beta$, orbital period $\tau$, satellite
scale radius $a_h$, satellite tidal radius $r_{\rm tid}$, satellite mass $M_{\rm Sgr}$, satellite spin orientation $S_{\rm Sgr}$ (0 = no spin, +1 = prograde
spin, $-1$ = retrograde spin), orbit orientation with respect to Galactic plane, Milky Way model (models MWa and MWb from \cite{WidDub}), satellite model
(MTL = mass traces light, 100\% MB = 100\% of most-bound particles are stars, 25\% MB = 25\% of most-bound particles are stars), satellite mass-to-light
ratio $M/L$.} \label{T:runs}
\end{center}
\end{table}

Having described our methodology, we can now take a look at the series of simulations performed.  Our strategy was to perform a default run with model
parameters that closely resemble those of the actual Sgr dwarf, then vary each of these parameters in turn to see just how robust asymmetric tidal streams
really are as a signature of dark-matter forces.  We have not yet developed a satisfactory method for inverting the final tidal stream into an initially
bound satellite, so we could not perform a rigorous comparison between our simulations and actual data or the simulations of \cite{2MassTT}.  Those
simulations used rigid potentials for the bulge, disk, and halo including a logarithmic as opposed to NFW halo potential, so we would not expect close
agreement in any case.  Nonetheless, for our default run we chose initial conditions that reproduced the best-fit values of \cite{2MassTT} as closely as
possible because this seemed at least a reasonable starting point.

All simulations were begun on an elliptical orbit at apocenter 59 kpc from the Galactic center, and were evolved for 2.4 Gyr (approximately 2.5 orbits).
The initial tangential velocity was chosen to achieve a pericenter distance of 14 kpc, the best-fit value of \cite{2MassTT} for models with spherical halos. 
Our default Galactic model (model MWb of \cite{WidDub}) was chosen to be consistent with observational constraints including the Galactic rotation curve and
the local velocity ellipsoid.  The Sgr halo normalization $\rho_h$ and scale radius $a_h$ of Eq.~(\ref{E:NFW}) were set by requiring the Sgr to virialize
with a concentration $c \equiv r_{\rm vir}/a_h$ of 15 at the start of the simulation.  The normalization and energy cutoff $E_h$ were then adjusted so that
the density profile was cut off at the tidal radius at apocenter and contained the desired final Sgr mass.  The total mass of stars and dark matter varied
between simulations, but we assumed throughout that the stars themselves had a $M/L$ ratio of 2.25 consistent with the purely stellar model of
\cite{HelWhi01}.  As the dark matter produces no light the total $M/L$ ratio of the satellite will always exceed this value.  Following \cite{2MassTT}, we
assumed that our Sgr stars obeyed the same density profile as the dark matter.  Although this assumption is undoubtably unrealistic, it allows for a more
direct comparison of the dark-matter and stellar tidal streams and is difficult to improve upon with current observations.  We will ultimately relax this
assumption in the simulations presented in subsection \ref{SS:2comp}.  A summary of the parameter values for all our simulations is given in
Table~\ref{T:runs}.

These parameter values, along with the particle numbers for each species and settings for GADGET-2, determine the initial conditions for our simulations.
In all the simulations presented here we used 10,000 particles each for the Galactic disk and bulge, 80,000 particles for the Galactic halo, and 200,000
particles for the satellite galaxy.  All satellite particles were assigned the same mass, with the relative number of particles representing stars and dark
matter determined by the mass-to-light ratio of the simulation.  The GADGET-2 settings were adjusted within the suggested ranges; in particular we chose
gravitational softening lengths $\epsilon$ of 0.05 kpc for the Galactic particles and 0.02 kpc for the satellite particles.  GADGET-2 uses a
gravitational-softening procedure that preserves the exact $1/r^2$ Newtonian force law for distances greater than $2.8\epsilon$ \cite{Gadget2}.  Doubling
the particle number and thus decreasing the ratio of inter-particle separation to softening length did not lead to significant changes in the morphology of
the tidal streams on the scales we are considering.  Most importantly, the asymmetry in the leading and trailing tidal streams we are proposing as a
signature of dark-matter forces was virtually unchanged (about 1\% of the deviation between models).  With the initial conditions established, we can
proceed to examine the results of the simulations.

\begin{figure}[t!]
\scalebox{0.90}{\includegraphics{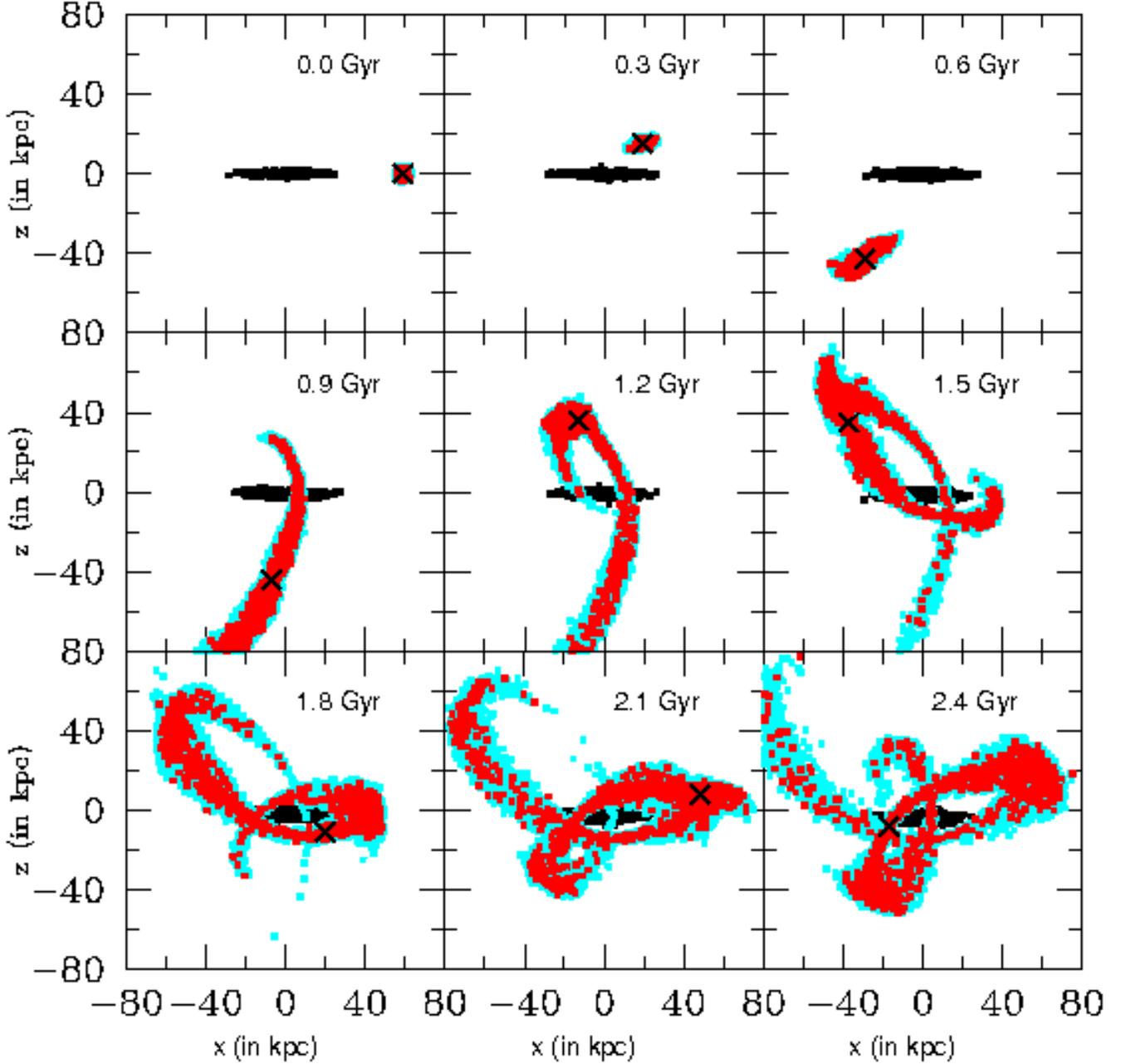}}
\caption{Tidal disruption of a satellite galaxy with mass and orbit similar to that of the Sgr dwarf.  Galactic disk particles are black, while the
satellite's dark matter and stars are shown in cyan and red (light and dark grey) respectively.  The top left panel shows the beginning of the simulation at
$t = 0.0$ Gyr, and the remaining panels show snapshots at 0.3 Gyr intervals going from left to right along each row as labeled.  The X's mark the location
of the satellite's bound core in each frame.  The satellite's orbit is counterclockwise about the Galactic center, so the leading (trailing) stream can be
identified by tracing counterclockwise (clockwise) from the location of the core.} \label{F:evo}
\end{figure}

First let us take a look at tidal disruption in our default $\beta = 0.0$ simulation, Run 1a of Table~\ref{T:runs}.  Snapshots of this simulation taken at
0.3 Gyr intervals are presented in Fig.~\ref{F:evo}.  The satellite's orbit, like that of the actual Sgr dwarf, is almost perpendicular to the Galactic disk
which is seen edge on in the $x-y$ plane of these figures.  The satellite begins at apocenter on a counterclockwise orbit with position $z = 0$ and
velocity in the $+z$ direction.  The initial conditions were chosen so the satellite just fills its Roche lobe at apocenter, implying that significant tidal
disruption isn't seen until $t = 0.6$ Gyr, the first snapshot after pericenter passage.  The satellite core is close to apocenter at $t = 0.6$ Gyr as
indicated by the small change in its position between this time and the next snapshot at $t = 0.9$ Gyr.  The variance in the velocities of the newly
disrupted stars and dark-matter particles has increased by $E_{\rm tid}$ however, and these particles are seen to disperse to a considerable extent between
the two snapshots.  The remaining panels show the continued development of the tidal streams, which by the final snapshot at $t = 2.4$ Gyr wrap around the
Galactic center alomst twice.

Multiple wrappings are an expected feature of tidal disruption, as elliptical orbits do not close for generic potentials including those of our Galactic
models.  The simulations of \cite{2MassTT} have been evolved for more orbits than our simulations and consequently show even more wrappings.  The SDSS
collaboration has identified two distinct peaks in their stellar luminosity function at certain positions on the sky, which they attribute to distinct
structures at different distances possibly corresponding to multiple wrappings of the Sgr stream \cite{SDSS}.  By $t = 2.4$ Gyr the streams in our
simulation reveal four different apocenters at about 2, 7, 10, and 12 o'clock with respect to the Galactic center.  Those at 7 and 12 o'clock belong to the
leading tidal stream, while the particles at 2 and 10 o'clock are part of the trailing stream.  The tidal material tends to accumulate at apocenter, where
potential energy is maximized and velocities are consequently smallest.  Note that apocenter distances in the trailing stream exceed the initial apocenter
distance of 59 kpc, while those in the leading stream are closer to the Galactic center.  This is exactly what one would expect if particles in the trailing
stream had gained energy during disruption while those in the leading stream had lost energy.  Since no dark-matter forces operate in Run 1a and the stars
and dark matter have the same initial distribution, the red points in Fig.~\ref{F:evo} are an unbiased subset of the cyan points.

\begin{figure}[t!]
\scalebox{0.90}{\includegraphics{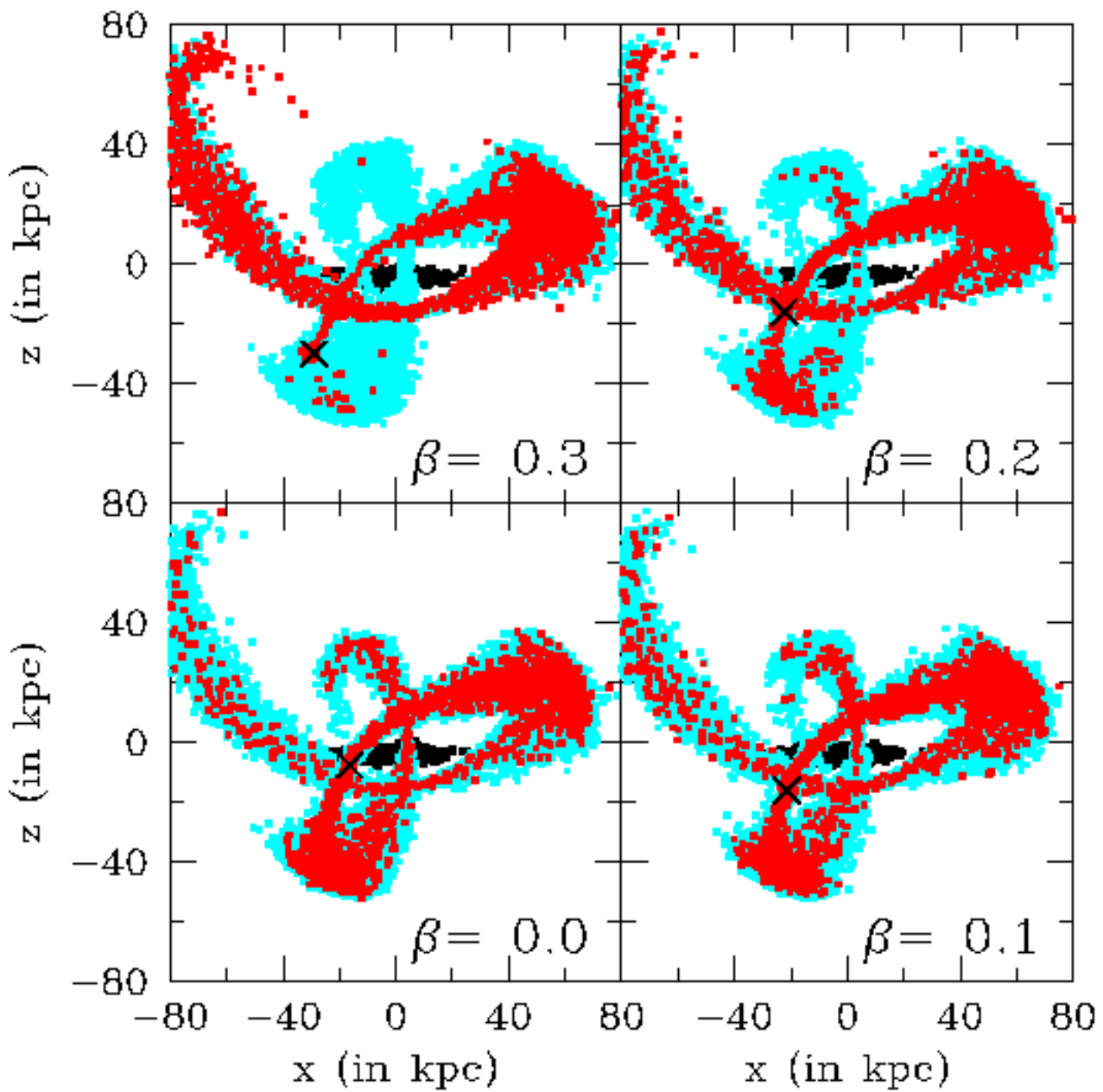}}
\caption{Tidal streams after 2.4 Gyr for four different values of $\beta$ corresponding to Runs 1a through 1d of Table~\ref{T:runs}.  The dark-matter force
increases in strength as labeled from $\beta = 0.0$ to $\beta = 0.3$ as one goes counterclockwise from bottom left to top left.  The bottom left panel
of this figure is identical to the bottom right panel of Fig.~\ref{F:evo}.} \label{F:beta}
\end{figure}

This is not the case in Fig.~\ref{F:beta}, where nonzero dark-matter forces lead to a redistribution of the stars (red particles) from the leading to the
trailing stream for reasons described in Section~\ref{S:tidal}.  Though the effect is barely noticeable for $\beta = 0.1$, a dark-matter force only 1\% the
strength of gravity, there is significant depletion in the leading stream for $\beta = 0.2$ and it is almost completely evacuated of stars for
$\beta = 0.3$.  The trailing stream is correspondingly enhanced, and we see that for $\beta = 0.3$ the very tail of the trailing stream appears to be
composed entirely of stars despite the fact that dark-matter particles are far more numerous overall.  This can perhaps be interpreted as evidence for our
supposition that after disruption stars would be left behind by $\pi \beta^2 f_{\rm sat} f_R$ radians per orbit compared to dark-matter particles on similar
orbits.  Though more detailed simulations of this kind would need to be compared directly to observed stellar densities along the stream to set any firm
limits on a dark-matter force, the relatively equal numbers of leading and trailing stars seen by 2MASS strongly suggests that a dark-matter force greater
than 10\% the strength of gravity is already excluded \cite{2MassTT}.

\begin{figure}[t!]
\scalebox{0.90}{\includegraphics{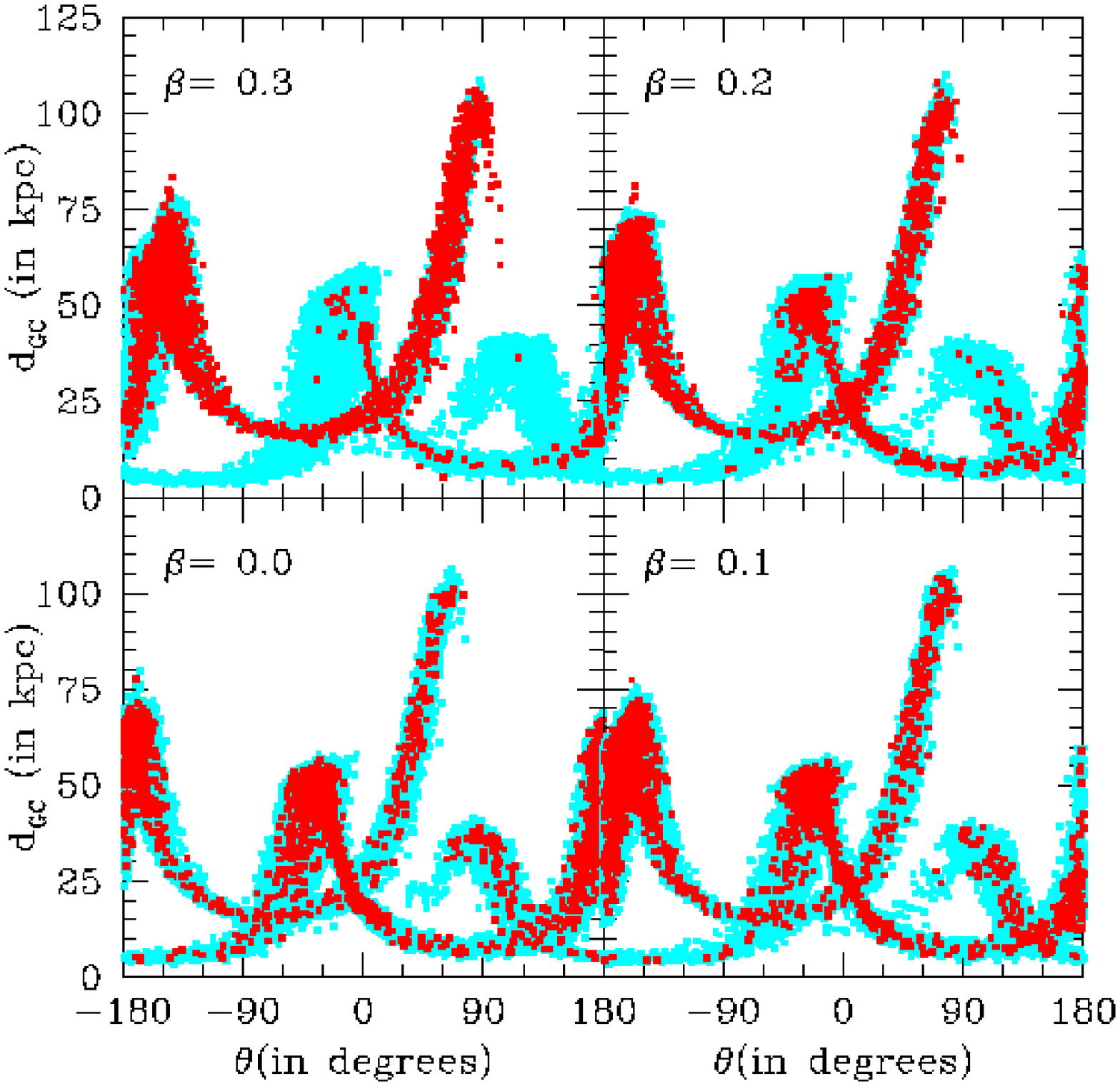}}
\caption{Galactocentric distances $d_{\rm GC}$ in kpc as a function of position $\theta$ along the tidal stream as viewed from the Galactic center.  The
four panels show simulations with different values of $\beta$ as labeled (Runs 1a through 1d), and correspond to the four panels of Fig.~\ref{F:beta}.
The satellite core is located at $\theta = 0$, while the trailing and leading streams begin in the positive and negative $\theta$ directions respectively.
} \label{F:betaD}
\end{figure}
\begin{figure}[t!]
\scalebox{0.90}{\includegraphics{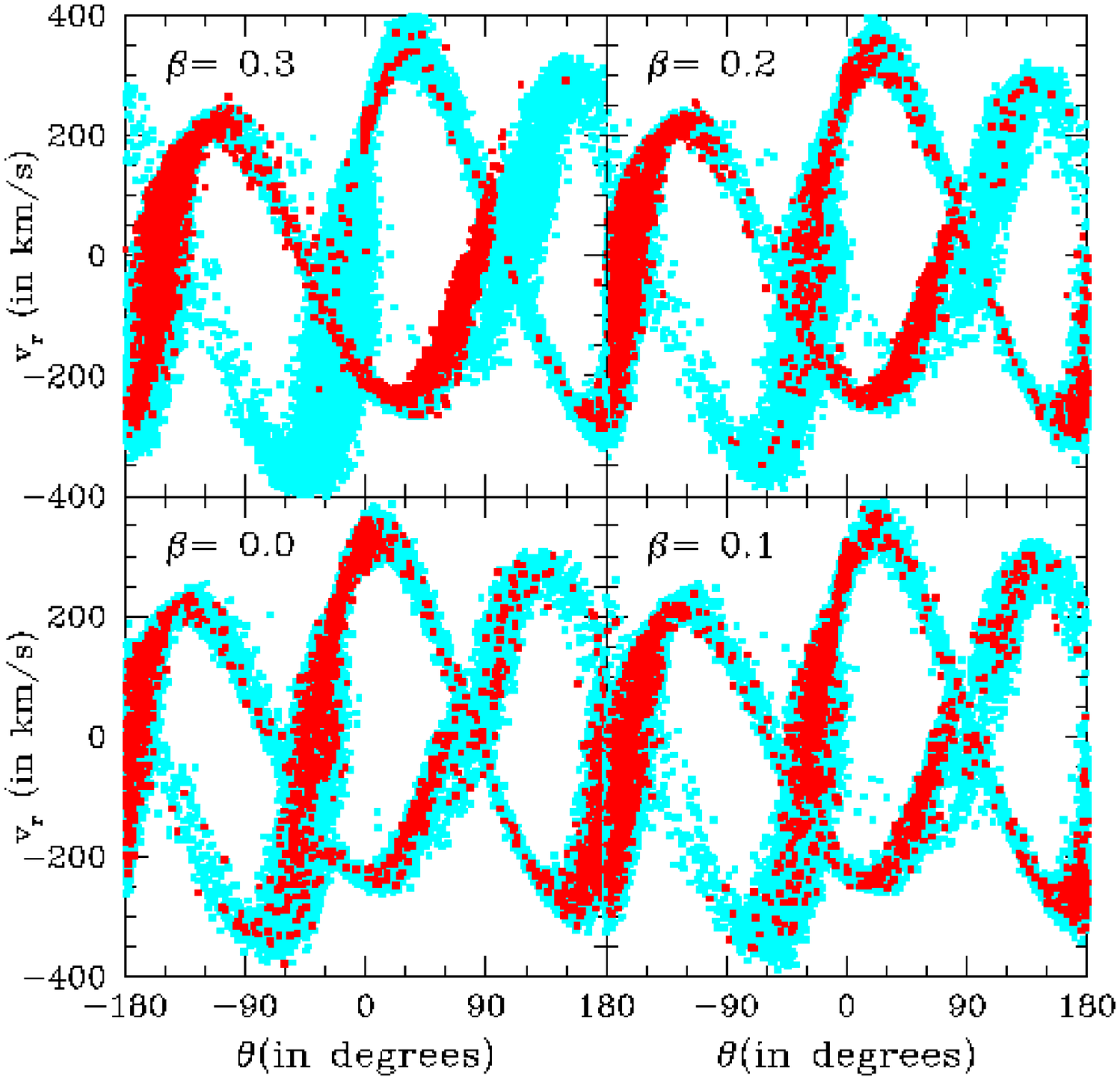}}
\caption{Radial velocities $v_{\rm r}$ in km/s as a function of position $\theta$ along the tidal stream as viewed from the Galactic center.  The four panels
show simulations with different values of $\beta$ as labeled (Runs 1a through 1d), and correspond to the four panels of Figs.~\ref{F:beta} and \ref{F:betaD}.
As previously, the satellite core is located at $\theta = 0$, while the trailing and leading streams begin in the positive and negative $\theta$ directions
respectively.} \label{F:betaV}
\end{figure}

Two more perspectives on these same simulations are shown in Figs.~\ref{F:betaD} and \ref{F:betaV}, where the Galactocentric distances and radial velocities
of the stars and dark-matter particles are shown as functions of $\theta$, the angular distance along the great circle formed by the satellite's orbit as
seen from the Galactic center.  Next to the angular position on the sky, these are the two easiest phase-space coordinates to measure, and they have already
been measured for a large sample of Sgr stream M-giants \cite{2MassObs,2MassVel}.  The satellite core is located at $\theta = 0$ in these figures, while the
trailing and leading streams begin in the positive and negative $\theta$ directions respectively.  Since the tidal streams wrap the Galaxy almost twice, the
two functions $d_{\rm GC}(\theta)$ and $v_r(\theta)$ are double valued over most of their domains.  The trailing stream runs off the right edge of the plots
at $+180^{\circ}$, then continues from the left edge at $-180^{\circ}$ to almost $+90^{\circ}$.  The leading stream similarly runs off the left edge of the
plots at $-180^{\circ}$, then continues from the right edge to beyond $0^{\circ}$ for the $\beta = 0.3$ case.  The four apocenters identified at 2, 7, 10,
and 12 o'clock in Fig.~\ref{F:beta} are seen in these figures in the intervals $(-180^{\circ}, -150^{\circ})$, $(-45^{\circ}, -15^{\circ})$,
$(+60^{\circ}, +90^{\circ})$, and $(+75^{\circ}, +105^{\circ})$ respectively.  Fig.~\ref{F:betaD} reveals that the four apocenter distances, beginning with
the edge of the leading stream and continuing to the tail of the trailing stream, are 40 kpc, 60 kpc, 75 kpc, and 110 kpc.  It is entirely expected that
apocenter distances should increase as one goes along the trailing stream, as the particles that lag the furthest behind are those that gained the most
energy and can therefore climb the highest in the Galactic potential well.  Dynamical friction may also play some role, though this should be minor as the
apocenter distance near the final position of the satellite core at 2.4 Gyr is close to the initial Galactocentric distance of 59 kpc.  Fig.~\ref{F:betaV}
shows that $v_r(\theta)$ is about $90^{\circ}$ out of phase with $d_{\rm GC}(\theta)$, as one would expect given that radial velocities vanish at
apocenter and pericenter, the extrema of Galactocentric distance.  Similar trends are seen in Figs.~10 and 12 of \cite{2MassTT}, where simulations of the
Sgr tidal streams are presented alongside 2MASS observations and those of other experiments.

\begin{figure}[t!]
\scalebox{0.90}{\includegraphics{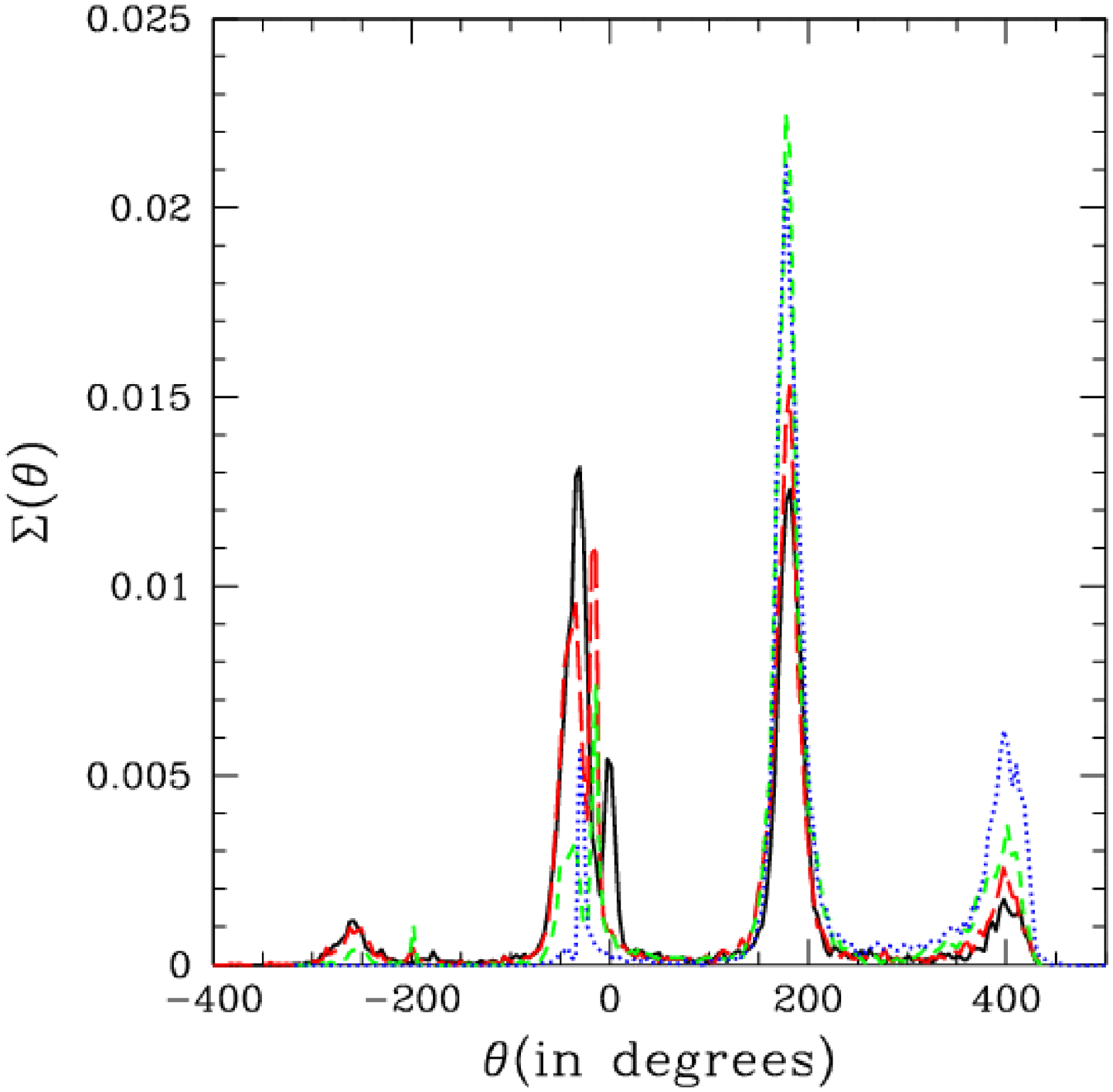}}
\caption{Surface density of stars as a function of angular distance $\theta$ along the tidal stream for Runs 1a through 1d.  As before, the satellite core
is located at $0^{\circ}$, while the trailing and leading streams are at positive and negative $\theta$ respectively.  This time however, the domain has
been expanded beyond $\pm 180^{\circ}$ to show the multiple wrappings of the stream side by side rather than on top of each other.  The four curves
correspond to the four panels of Fig.~\ref{F:beta}, with black (solid), red (long-dashed), green (short-dashed), and blue (dotted) curves belonging to the
$\beta = 0.0$ through 0.3 simulations.} \label{F:SD1}
\end{figure}

As we have predicted qualitatively and now seen in simulations, an attractive dark-matter force reduces the number of stars in leading tidal streams and
enhances stellar densities in the trailing streams.  A normalized count $\Sigma(\theta)$ of the stars per radian along the stream length for Runs 1a through
1d is presented in Fig.~\ref{F:SD1}, where in contrast to Figs.~\ref{F:betaD} and \ref{F:betaV} we have expanded the domain beyond $\pm 180^{\circ}$ to fully
``unwind'' the stream from the Galactic center.  Though this is trivial to do in simulations where we can track the trajectories of individual particles,
it is much more difficult to do with real observations where a true leading star cannot readily be distinguished from a star that trails behind by almost a
full orbit.  Fig.~\ref{F:evo} shows that the tidal stream crosses itself after 1.5 Gyr, and it is not obvious from this plot how to identify leading and
trailing stars in the vicinity of the intersection.  The key to accomplishing this identification is to include radial-velocity information, as the stream
intersections occur at different values of $\theta$ in Figs.~\ref{F:betaD} and \ref{F:betaV}.  This implies that no intersections exist in the
three-dimensional phase formed by the observationally accessible angular position $\theta$, Galactocentric distance $d_{\rm GC}$, and radial velocity $v_r$.
We have found that by tracing maximum-density contours through this three-dimensional phase space, we can differentiate leading and trailing stars with high
accuracy.  The ability of real experiments to differentiate stars in this manner will depend on instrumental resolution and other systematics, and will need
to be investigated further before such techniques can be applied in practice.  We are hopeful that leading and trailing stars can be distinguished to some
degree by current experiments, and with great reliability by future astrometry missions that will have proper motions as well.  Fig.~\ref{F:SD1} was
prepared using this method and should be interpreted as an optimistic but reasonable estimate of future experimental possibilities.

The four peaks seen in Fig.~\ref{F:SD1} correspond to the four apocenter passages, because as previously mentioned particles tend to accumulate in these
positions where velocities are minimized.  The decrease in the two leading peaks at $-250^{\circ}$ and $-50^{\circ}$ with increasing $\beta$ is apparent,
as is the corresponding increase in the trailing peaks at $+175^{\circ}$ and $+400^{\circ}$.  The ratio of the number of stars in the leading peak at
$-250^{\circ}$ to those in the trailing peak at $+400^{\circ}$ can serve as a crude measure of the effects of a dark-matter force.  This leading-to-trailing
ratio drops from 0.66 in the absence of a dark-matter force to 0.44, 0.091, and 0.0042 for $\beta = 0.1$, 0.2, and 0.3 respectively.  The SDSS has observed
hundreds of stars per square degree in the Sgr tidal stream \cite{SDSS}, so the {\it statistical} errors associated with measurements of the
leading-to-trailing ratio should be small.  The real uncertainty lies in an unforeseen {\it systematic} bias in observing stars in one direction on the
sky compared to another.  It is also important to verify that this signature is robust to changes in our Galactic and Sgr models, and not degenerate with a
more conventional change to the system than the addition of a dark-matter force.  It was with this object in mind that we undertook Runs 2 though 9 in
Table~\ref{T:runs}, which we examine in turn in the remainder of this section.

\subsection{Satellite Mass} \label{SS:mass}

\begin{figure}[t!]
\scalebox{0.90}{\includegraphics{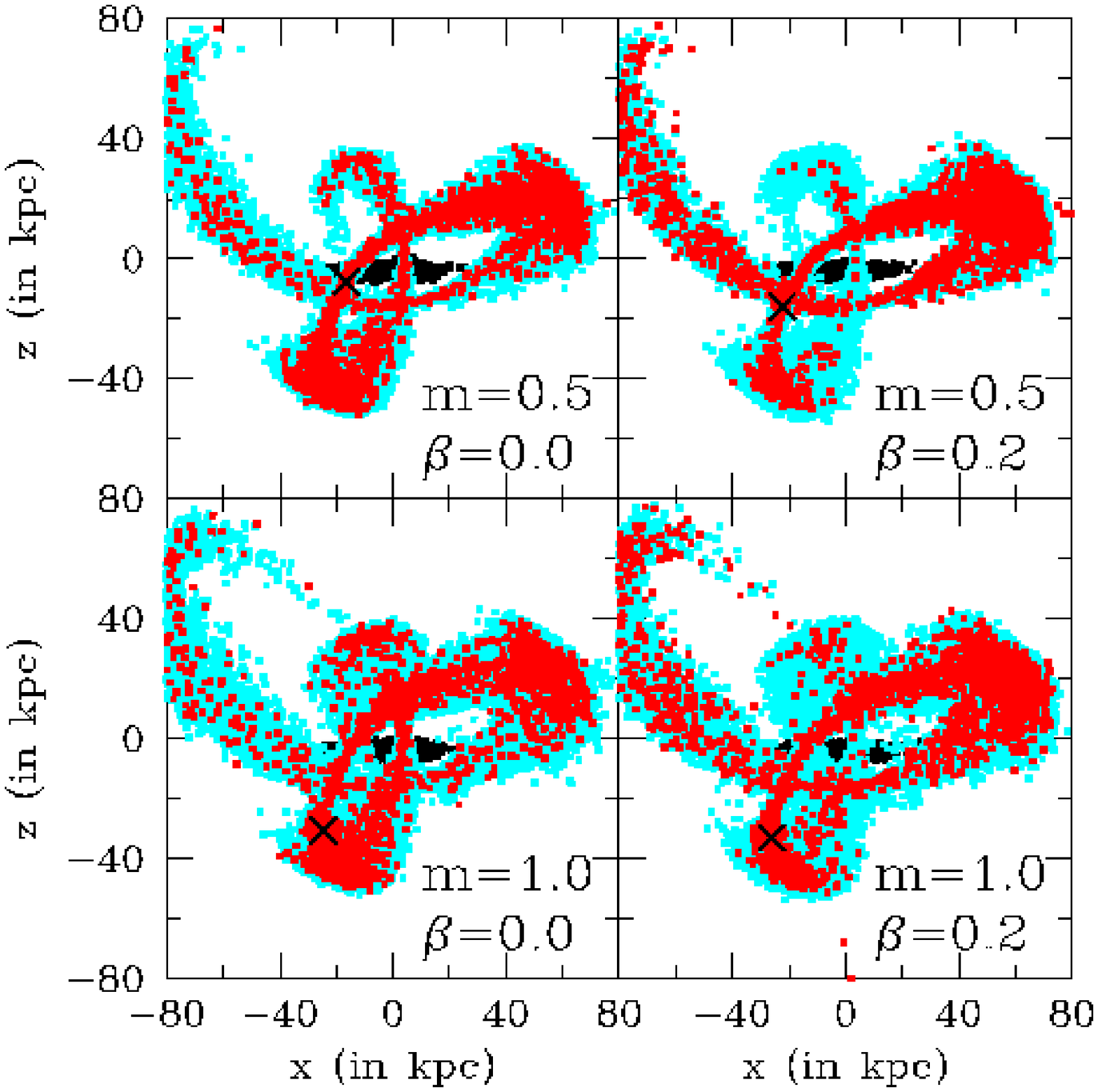}}
\caption{Tidal disruption for satellite galaxies of initial masses of 0.5 and 1.0 as measured in units of $10^9 M_{\odot}$.  The left panels
(Run 1a, top and Run 2a, bottom) have no dark-matter force while the right panels (Run 1c, top and Run 2c, bottom) have $\beta = 0.2$ as labeled.}
\label{F:mass}
\end{figure}

One obvious parameter in our models to change is the initial satellite mass, which was chosen to be $5.0 \times 10^8 M_{\odot}$ in the simulations presented
so far.  This is at the upper bound of the range $M_{\rm Sgr} = (2-5) \times 10^8 M_{\odot}$ of possible {\it final} masses of the Sgr dwarf as
determined by velocity-dispersion measurements \cite{2MassTT}.  However the Sgr dwarf could easily have lost half of its mass to tidal disruption, so an
initial mass of $10^9 M_{\odot}$ might more accurately reflect the upper bound of possible masses.  We used this initial mass in Runs 2a through 2c as listed
in Table~\ref{T:runs}, which led to correspondingly larger scale radii $a_h$ and tidal radii $r_{\rm tid}$.  Runs 2a and 2c are depicted in
Fig.~\ref{F:mass}, along with the $5.0 \times 10^8 M_{\odot}$ Runs 1a and 1c for comparison.  The most obvious effect of the increased satellite mass is an
increase in the thickness of the tidal streams, which is expected as the debris width reflects the satellite's binding energy $E_{\rm bin}$ at the time of
disruption \cite{2MassTT}.  Since the tidal radius increases, the typical energy $E_{\rm tid}$ imparted during tidal disruption increases as well.  This
leads to {\it longer} tidal streams which are particularly noticeable in the trailing streams.  In fact, the two stellar particles visible near $(0, -80)$
in the bottom right panel of Fig.~\ref{F:mass} belong to the stellar tail of the trailing tidal stream which now extends clear across the Galaxy from its
previous position.  The particles in this and other figures have been downsampled for presentation purposes, so these two particles do represent a larger,
statistically significant population.  For the purposes of this paper, the importance of these simulations is that the increase in satellite mass increases
the amount of tidal debris in {\it both} the leading and trailing streams, and that an attractive dark-matter force still leads to a pronounced drop in the
leading-to-trailing ratio.

A surprising feature of these simulations that warrants further comment is that the leading streams of the $10^9 M_{\odot}$ runs appear to be more enhanced
than the trailing streams, leading to greater overall symmetry between the two streams.  This is contrary to our expectation based on the analytical
argument of Section~\ref{S:tidal} that the natural asymmetry $\Delta r_{\rm nat}$ should increase with satellite mass.  The probable explanation for this is
dynamical friction, a factor that did not enter into that order-of-magnitude estimate.  Dynamical friction will be greater for the more massive satellite
\cite{B&T}, implying that it will fall deeper into the Galactic potential and advance further along its orbit as indicated by the relative positions of the
crosses in the two left panels of Fig.~\ref{F:mass}.  This effectively provides more time for the tidal debris to advance into the foremost leading-stream
apocenter and retreat from the rearmost trailing-stream apocenter.  It is possible that a comparison between to the two simulations at equal orbital phases
rather than equal elapsed times would have been more appropriate, but this might have introduced additional discrepancies.  This issue will need to be
explored in future work when this model is confronted with data and elapsed time is a parameter to be varied.

\subsection{Satellite Spin} \label{SS:spin}

\begin{figure}[t!]
\scalebox{0.90}{\includegraphics{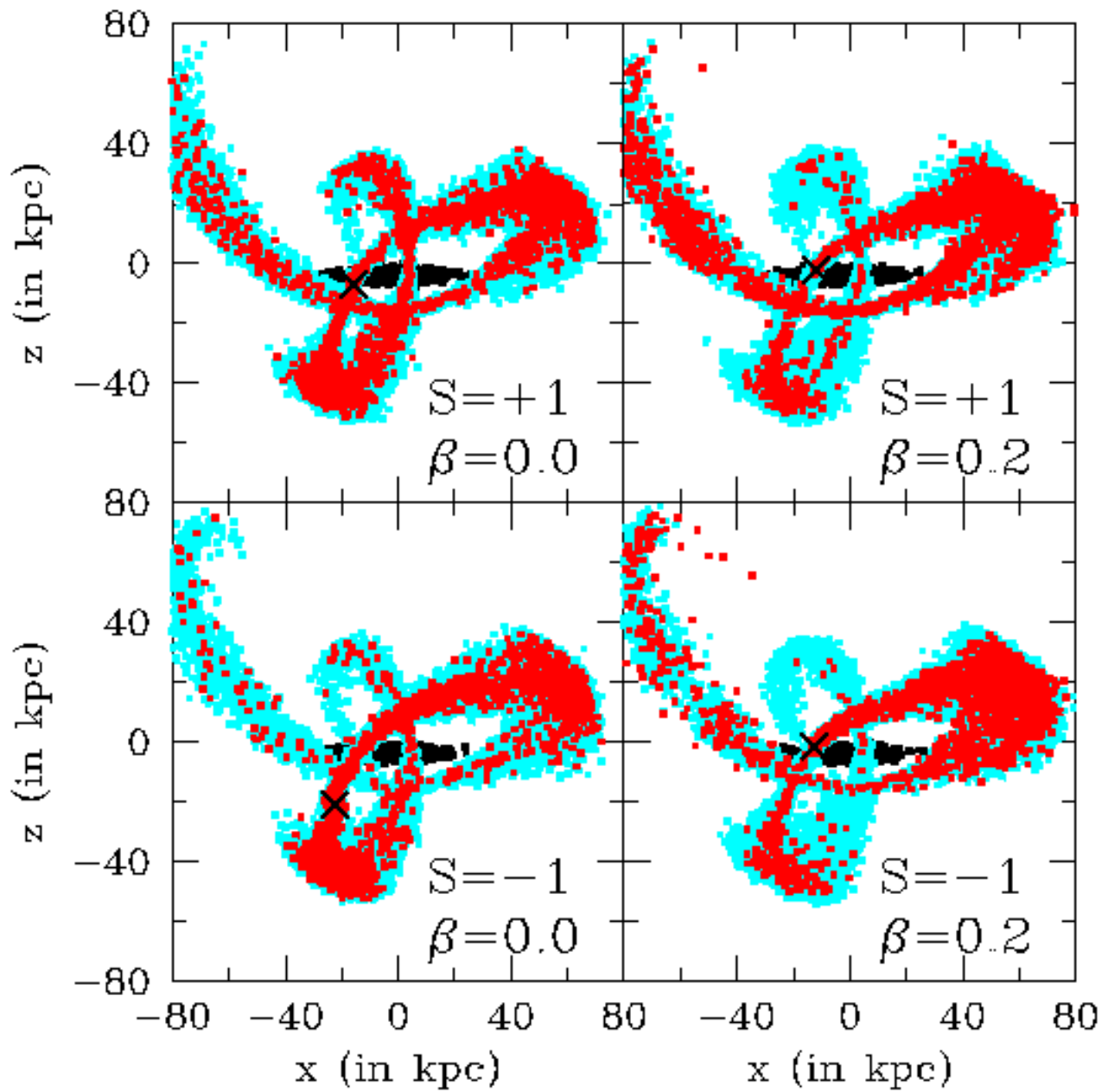}}
\caption{Tidal disruption for satellite galaxies with maximal prograde and retrograde angular momentum.  The top two panels $(S = +1)$ have prograde angular
momenta (parallel to the orbital angular momenta).  The bottom two panels $(S = -1)$ have retrograde angular momenta (anti-parallel to the orbital angular
momenta).  As in Fig.~\ref{F:mass}, the left panels have no dark-matter force while the right panels have $\beta = 0.2$.} \label{F:spin}
\end{figure}

Another possibility to consider is whether the satellite galaxy has any net rotation, which we will call spin angular momentum to distinguish it from the
orbital angular momentum of the satellite about the Galactic center.    As discussed in Section~\ref{S:sim}, our systems are constructed from distribution
functions of $E$ and $L_z$.  The density profile $\rho(r)$ is determined exclusively by the part of the distribution function {\it even} in $L_z$,
while the mean streaming velocity $\bar{v_{\phi}}$ is determined by the part {\it odd} in $L_z$ \cite{B&T}.  While this criterion would seem to suggest
that the value of the total spin angular momentum is independent of the density profile, the requirement that the distribution function be everywhere
positive definite sets a maximum value for the spin.  This maxmimum value depends on our choice of distribution function for the satellite, which was a
fourth-order polynomial in $E$ that closely approximates an NFW profile \cite{WidNFW}.  By splitting this function into parts with positive and negative
$L_z$ and weighting them by $(1 \pm S)/2$, where $S$ is the spin parameter given in Table~\ref{T:runs}, we can vary the spin angular momentum within our
model.  In terms of the dimensionless spin parameter \cite{Peacock},
\begin{equation}
\lambda \equiv \frac{L |E|^{1/2}}{GM^{5/2}} \, ,
\end{equation}
our non-rotating default simulations have $\lambda < 10^{-3}$ while the maximally rotating simulations presented in this subsection have
$\lambda \simeq 0.11$.  This is somewhat larger than the mean value $\lambda \simeq 0.05$ expected in pressure-supported systems \cite{Peacock} but is
physically reasonable for both the dark matter and stars.  There is also no obvious astrophysical reason to expect an alignment between the
spin and orbital angular momenta, but one would expect any possible coupling between them to be maximized in this case.  We therefore investigate cases
where the spin is 100\% prograde (parallel to the orbital angular momenta) or 100\% retrograde (anti-parallel).

Simulations 3a through 3d, described in Table~\ref{T:runs} and illustrated in Fig.~\ref{F:spin}, involve satellites with maximal spin angular momentum.  We
see that prograde spin significantly enhances tidal disruption, while retrograde spin inhibits disruption.  This can be understood in the context of our
arguments concerning tidal disruption in Section~\ref{S:tidal}.  For a satellite on a circular orbit, stars disrupted from the far side of the satellite
are higher in the Galactic potential well.  They therefore tend to gain energy with respect to the satellite and go on to form the trailing tidal stream.
This tendency is reinforced for stars on prograde orbits, as the spin and orbital velocities are parallel on the far side of the satellite and therefore
add coherently.  Conversely, stars disrupted from the near side of the satellite are deeper in the Galactic potential well and lose energy to form a leading
stream.  The tendency to lose energy is again fostered by being on a prograde orbit, as on the near side of the satellite the spin and orbital velocities
are anti-parallel and partially cancel.  Stars on retrograde orbits are more likely to remain bound to the satellite, as their spin velocities will cancel
with the orbital velocities on the far side of the satellite and add coherently on the near side.  Spin effectively leads to a smaller (larger) tidal radius
for particles on prograde (retrograde) orbits \cite{Read}.  One wonders if the preferential disruption of stars on
prograde orbits will produce an observable net retrograde rotation in the remaining bound stars of highly disrupted systems.  For satellites on nearly
radial orbits about the Galactic center, stars on prograde and retrograde orbits will be disrupted in equal numbers as the spin velocities on the near and
far sides of the satellite are orthogonal to the orbital velocities.  This analysis and the simulations shown in Fig.~\ref{F:spin} thus reveal that net
rotation can play a major role in tidal disruption, but it does not produce a marked asymmetry in the leading and trailing streams like a dark-matter force.
While the top right and bottom left panels of Fig.~\ref{F:spin} have quite similar leading streams, retrograde orbits in the bottom left panel have reduced
stellar densities in the trailing stream as well.  In contrast, the trailing stream in the top right panel has been substantially enhanced by a dark-matter
force.  Comparisons of the leading and trailing streams are essential for a claimed detection of a dark-matter force, and a measured rotation curve for the
satellite would be enormously useful as well.  The Sgr dwarf is observed to have little net rotation \cite{SgrRot}, so we do not expect spin to have a
substantial effect on its tidal streams.

\subsection{Circular Orbit} \label{SS:circ}

\begin{figure}[t!]
\scalebox{0.90}{\includegraphics{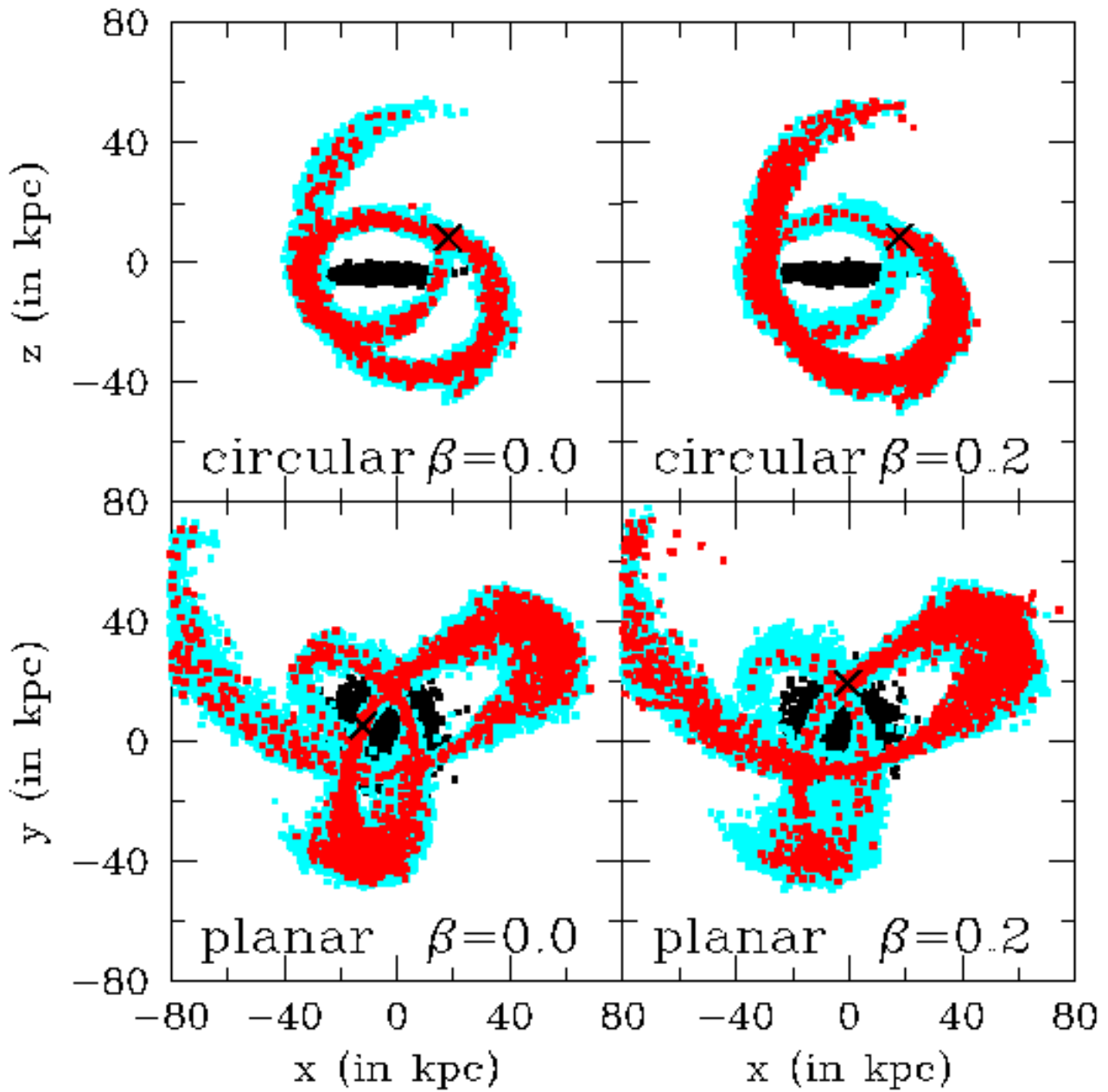}}
\caption{Tidal disruption for satellite galaxies on orbits significantly different from that of the Sgr dwarf.  The top panels show a satellite on an orbit
with a period similar to that of the Sgr dwarf, but much lower eccentricity.  The orbit of the satellite in the bottom panels has a period and eccentricity
similar to that of the Sgr dwarf, but lies in the Galactic plane.  The Sgr dwarf is on an almost polar orbit.  As in Fig.~\ref{F:mass}, the left panels have
no dark-matter force while the right panels have $\beta = 0.2$.} \label{F:orbit}
\end{figure}

Having examined changes to the satellite galaxy itself, we now look at how changes in the satellite's orbit might affect the use of asymmetric tidal streams
as a probe of a dark-matter force.  One possible change is to reduce the orbit eccentricity, which we did in Runs 4a and 4b described in Table~\ref{T:runs}.
The initial Galactocentric distance was reduced as well, to maintain the same semi-major axis and orbital period of our previous simulations.  Allowing the
same number of orbits over the course of the 2.4 Gyr simulations allows for roughly equal development of the tidal streams.  However, conserving the
semi-major axis with minimal eccentricity requires us to increase the pericenter distance as well.  In previous simulations, the bulk of the tidal
disruption occurred at pericenter, so one might expect the simulations with nearly circular orbits to exhibit reduced tidal disruption.  This is indeed seen
in the top panels of Fig.~\ref{F:orbit}, where the tidal streams are noticeably thinner than those in Fig.~\ref{F:beta}.  We used the same satellite models
in Runs 4a and 4b as we did in Runs 1a and 1c, though the reduced initial Galactocentric distance implied that the satellites would be overflowing their
Roche lobes from the beginning of the simulations.  More realistic initial conditions would have taken into account that a satellite on a circular orbit had
slowly spiralled inwards due to dynamical friction, and would have cut off the density profile at the tidal radius.  However such a satellite would have
been much smaller than that in Runs 1a and 1c making direct comparisons more difficult.  Despite the excess material initially outside the tidal radius, the
simulations of Fig.~\ref{F:orbit} display the reduced tidal disruption we expect due to the larger pericenter distance.

Fig.~\ref{F:orbit} also reveals another complication of circular orbits, the possibility of significant overlap between the leading and trailing streams.
Although as before the trailing stream is at larger Galactocentric distances than the leading stream, this separation is much less than that for a satellite
on a highly eccentric orbit.  Eccentric orbits in the combined potential of the bulge-disk-halo system do not close, so it is comparatively easy to trace
from the edge of the leading stream to the tip of the tidal tail.  Circular orbits in an axisymmetric potential will close if they lie in the symmetry plane
or are perfectly orthogonal to it as ours is.  Fig.~\ref{F:orbit} shows that the leading and trailing streams consequently overlap to a considerable extent
near 9 o'clock.  Although we do see significant depletion (enhancement) of stars in the leading (trailing) stream where there is no overlap, the difficulty
in separating the leading and trailing streams in the overlap region could pose problems to practical efforts to determine the leading-to-trailing ratio.

\subsection{Planar Orbit} \label{SS:plane}

Identifying stream stars may also be a problem when the satellite's orbit lies in the Galactic plane, as in Runs 5a and 5c shown in the bottom panels of
Fig.~\ref{F:orbit}.  We are fortunate that the orbital poles of the Sgr dwarf at Galactic latitudes $b = \pm 13^{\circ}$ are almost orthogonal to the
Galactic poles.  Fig.~\ref{F:orbit} shows significant overlap between the black Galactic disk and the colored inner portions of the tidal streams.  This
figure understates the true extent of the problem, as the Galactic disk was simulated with many fewer particles than the tidal streams since we were less
concerned with resolving its features.

Another reason to perform the planar simulations of Fig.~\ref{F:orbit} was that the combined bulge-disk-halo system, while axisymmetric, is nonspherical.
Perhaps orbits and thus the tidal streams might look different in the plane of symmetry.  Comparing the bottom two panels of Fig.~\ref{F:orbit} to the
panels of Fig.~\ref{F:beta} with the same values of $\beta$, we find that this is not the case.  Other than a slight counterclockwise rotation in the
longitude of pericenter, the qualitative features of the tidal streams appear the same including the leading-to-trailing asymmetry in the presence of a
dark-matter force.

\subsection{Milky Way Model} \label{SS:MW}

\begin{figure}
\scalebox{0.90}{\includegraphics{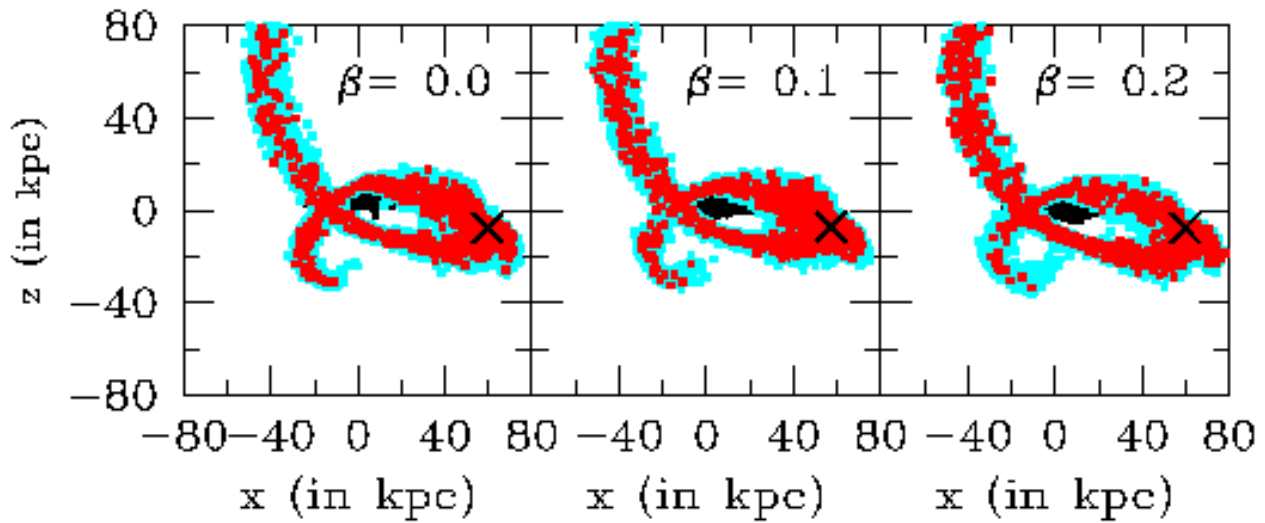}}
\caption{Tidal disruption of satellite galaxies in an alternative model of the host galaxy, MWa of \cite{WidDub}.  Dark-matter forces increase in strength
from left to right as labeled.} \label{F:MWa}
\end{figure}

\begin{table}
\begin{center}
\begin{tabular}{ | c | c | c | c | c | c |}
\hline
Model & $M_b$ (in $ 10^{10} M_{\odot}$) & $M_d$ (in $ 10^{10} M_{\odot}$) & $M_h$ (in $ 10^{10} M_{\odot}$) & $a_h$ (in kpc) & $M_h$(36.5 kpc)
(in $ 10^{10} M_{\odot}$) \\ \hline
MWa   & 1.20				& 4.65				  & 73.8			    & 12.96	     & 10.7	\\ \hline
MWb   & 1.19				& 3.52				  & 71.6			    & 8.818	     & 16.5	\\ \hline
\end{tabular}
\caption{Parameters for the two models of the Milky Way used in our simulations.  These are the bulge mass $M_b$, disk mass $M_d$, halo mass $M_h$, halo
scale radius $a_h$, and halo mass within 36.5 kpc, the semi-major axis of the Sgr orbit.} \label{T:MWmodels}
\end{center}
\end{table}

While planar and polar orbits in our model of the Milky Way may be similar, there is still significant freedom to adjust the model itself.  In simulations
6a through 6c, we explore tidal disruption using a very different model of the Milky Way (model MWa of \cite{WidDub}) that is still consistent with
observations like the Galactic rotation curve.  Parameters for this model and our default model MWb are shown in Table~\ref{T:MWmodels}.  Although the new
model MWa has a slightly more massive dark-matter halo, the larger scale radius implies that there is less mass within 36.5 kpc, the semi-major axis of the
Sgr orbit.  This leads to a longer orbital period and larger tidal radius as listed in Table~\ref{T:runs}.  The tidal streams shown in Fig.~\ref{F:MWa}
therefore wrap around the Galactic center fewer times than those Fig.~\ref{F:beta} in which model MWb was used.  Nonetheless, as in previous simulations, an
attractive dark-matter force leads to a pronounced decrease in the leading-to-trailing ratio of stars.  This can be seen in Fig.~\ref{F:MWa} by comparing
the leading stream at 7 o'clock to the long trailing stream at 11 o'clock.

\subsection{Two-Component Satellite} \label{SS:2comp}

\begin{figure}[t!]
\scalebox{0.90}{\includegraphics{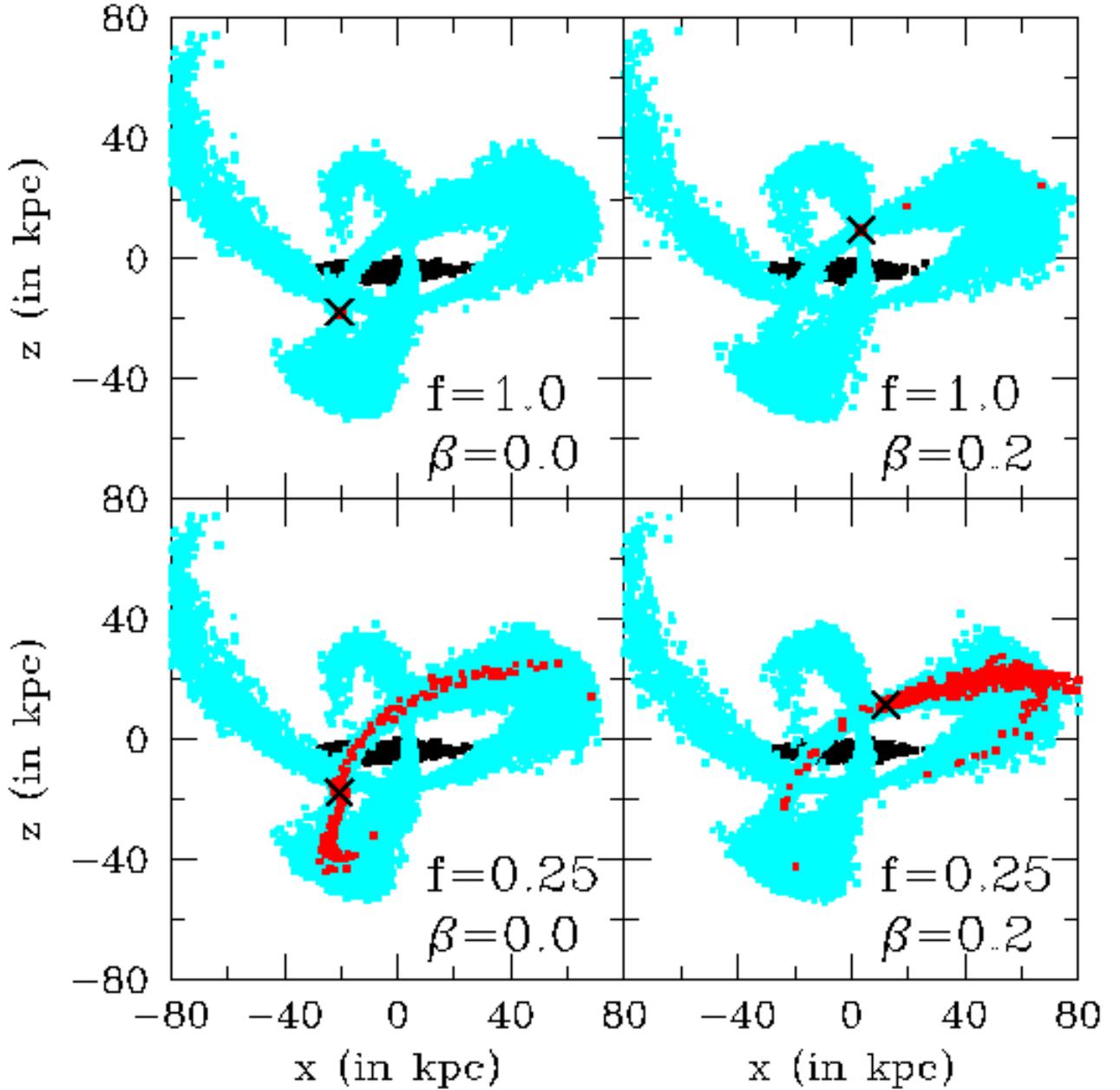}}
\caption{Tidal disruption of satellite galaxies in which the light no longer traces the mass as in previous simulations.  Although the initial density
distribution is still a modified NFW profile, the $n_{\ast}$ red stellar particles now represent a fraction $f$ of the $n_{\ast}/f$ most tightly bound
particles.  As labeled, the top panels have $f = 1.0$ while the bottom panels have $f = 0.25$.  The satellite cores are clearly visible in the top panels,
and are located in nearly the same positions in the bottom panels.  As in several previous figures, the left panels have no dark-matter force while the
right panels have $\beta = 0.2$.} \label{F:2C}
\end{figure}

If there is freedom to modify the model of the Milky Way consistent with current observations, there is even more freedom to modify our model of the
satellite galaxy.  When we see galaxies in isolation such as the Milky Way or M31, the stars are more centrally concentrated than the dark matter as the gas
was able to cool prior to star formation.  It is therefore inaccurate to assume that mass traces light on galactic scales as we have done in our model of the
satellite.  However, even at the beginning of our simulations, the satellite is not in isolation but presumably has been orbiting the Galactic center for at
least several Gyr.  Dark matter in an extended halo has already been stripped away, and by the time a stellar tidal stream begins to form the stars like the
dark matter must extend out to the tidal radius.  Although the detailed density profiles of the stars and dark matter may still be different, the assumption
that mass traces light is no longer quite as unreasonable.  Another excuse for making this assumption is that observations of the Sgr dwarf are not yet
capable of determining distinct profiles for the two components.  The 2MASS collaboration assumes that mass traces light in all their simulations of the Sgr
tidal streams \cite{2MassTT}.

Despite these arguments for a satellite in which mass traces light, we would like to make sure that this assumption does not artificially affect the
tidal-stream asymmetry induced by a dark-matter force.  To test this, we kept the modified NFW profile used previously for the satellite density
distribution but altered our choice of which particles represented stars and which represented dark matter.  The total mass-to-light ratio was preserved;
only the distributions of the stars and dark matter were changed.  In runs 7a and 7b we chose the $n_{\ast}$ most tightly bound particles to represent the
stars, while in runs 7c and 7d a randomly chosen 25\% of the $4n_{\ast}$ most bound particles represent stars.  Tidal disruption with these initial
conditions is shown in Fig.~\ref{F:2C}.  The top left panel, illustrating run 7a, shows that without a dark-matter force none of the most tightly bound
particles are disrupted.  Even a dark-matter force with $\beta = 0.2$ only manages to produce a weak trailing stream composed of a handful of particles.
The bottom two panels, depicting runs 7c and 7d, reveal that a somewhat more loosely bound distribution of stars does form both leading and trailing
streams, but much less extensive than those in Fig.~\ref{F:beta} where mass traces the light.  If real observational data is used to constrain a dark-matter
force, the total number of stars in the streams will be fixed; what matters is whether a two-component distribution could somehow mask the asymmetric tidal
tails we expect.  Fig.~\ref{F:2C} suggests at least qualitatively that this is not the case.  Simulations of the tidal disruption of two-component satellites
in previous work showed that there was no pronounced asymmetry in the tidal streams in the absence of a dark-matter force \cite{2Csat}.

\subsection{Satellite Mass-to-Light Ratio}

\begin{figure}[t!]
\scalebox{0.90}{\includegraphics{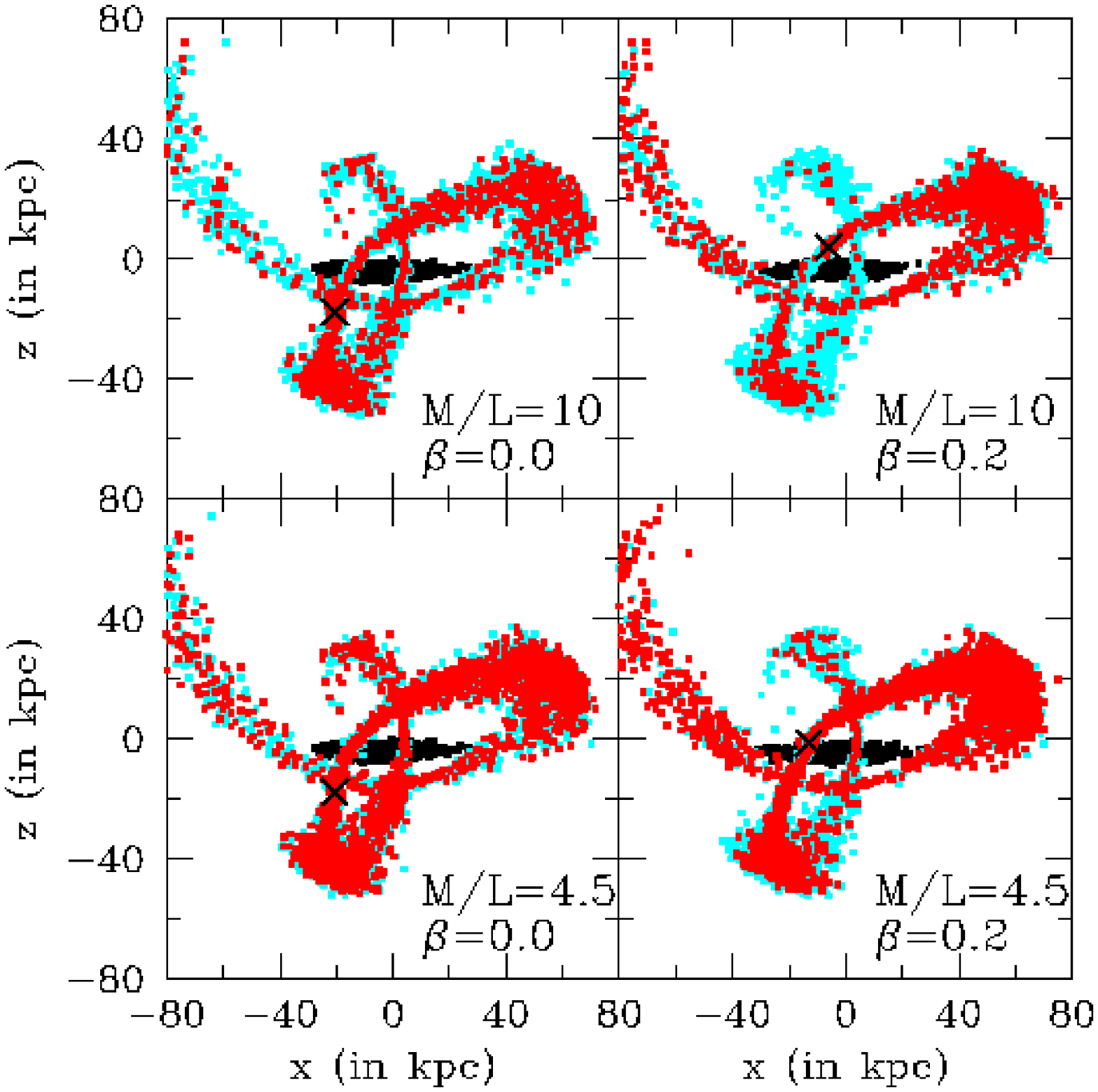}}
\caption{Tidal disruption of satellite galaxies with mass-to-light $(M/L)$ ratios of 10 and 4.5 in units of $M_{\odot}/L_{\odot}$, lower than $M/L = 40$ in
previous simulations.  With more stars and less dark matter, these satellites exhibit less of a tidal stream asymmetry for fixed stellar densities and
dark-matter force.  As in previous figures, the left panels have no dark-matter force while the right panels have $\beta = 0.2$.} \label{F:MLR}
\end{figure}

One factor that will mask evidence of a dark-matter force is the absence of dark matter from a satellite.  Objects such as globular clusters are not
expected to have dynamically significant amounts of dark matter, and therefore the asymmetry of their leading and trailing tidal streams should be
insensitive to the value of $\beta$.  However as long as the satellite is dark-matter dominated, its tidal streams will not depend strongly on the precise
fraction of the mass in stars.  The top panels of Fig.~\ref{F:MLR} show the tidal streams of a satellite with a mass-to-light ratio of 10 (measured in units
of $M_{\odot}/L_{\odot}$), implying that it has about four times as many stars as in previous simulations.  This is below the lower bound on the
mass-to-light ratio of the Sgr dwarf set by 2MASS data \cite{2MassTT}, assuming a $M/L$ ratio of 2.25 for the stellar population as in the purely stellar
model of \cite{HelWhi01}.  Despite the higher mass fraction of stars, the leading-to-trailing ratio of these simulations (Runs 8a and 8b) does not differ
significantly from that in the comparable default simulations (Runs 1a and 1c).

A more extreme case of a high stellar mass fraction is seen in the bottom panels of Fig.~\ref{F:MLR}, where the $M/L$ ratio of 4.5 implies that the satellite
is now 50\% stars by mass.  The dynamics of the bound core of the satellite is no longer totally driven by the dark matter, and as such the effects of a
dark-matter force on tidal disruption are less pronounced.  While the bottom right panel of Fig.~\ref{F:MLR} shows a greater total number of stars in the
tidal streams, the leading-to-trailing ratio is higher than in any other simulation with $\beta = 0.2$.  Even in this case, the leading-to-trailing ratio is
significantly below that of any run without a dark-matter force implying that a detection is not infeasible in such a system.

\subsection{Repulsive Dark-matter Force} \label{SS:rep}

\begin{figure}
\scalebox{0.90}{\includegraphics{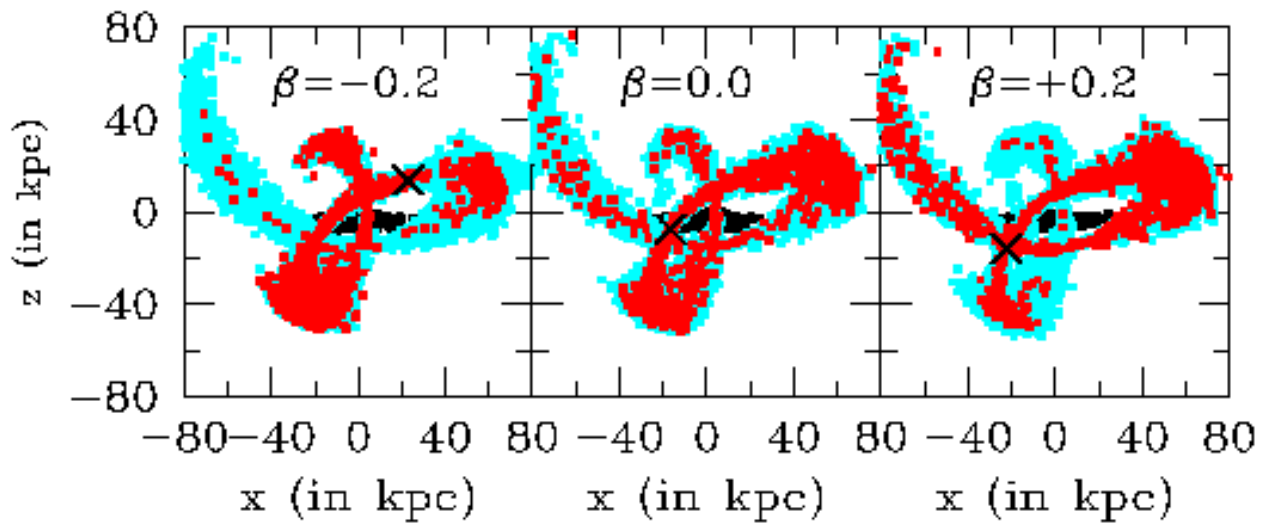}}
\caption{Tidal disruption of satellite galaxies in the presence of both attractive and repulsive dark-matter forces.  In the left panel, the satellite dark
matter has a charge-to-mass ratio $\beta = -0.2$ while the dark matter of the host galaxy's halo has $\beta = 0.2$.  The dark-matter force is therefore
attractive within each halo but repulsive between halos.  For comparison, the center panel has no dark-matter forces and the right panel has $\beta = 0.2$
for all dark-matter particles, leading to purely attractive dark-matter forces.} \label{F:neg}
\end{figure}

A final consideration is whether tidal streams can probe repulsive dark-matter forces as effectively as the attractive forces we have studied to this point.
While such forces are not permitted by the Lagrangian of Eq.~(\ref{E:L1}) that includes a single dark-matter species, they are allowed in models like those
of \cite{string,STcosmo} with multiple species.  Forces mediated by scalar particles are attractive for dark-matter particles of like charge, and repulsive
for particles of opposite charge.  In Fig.~\ref{F:neg}, we present Run 9 in which the satellite dark-matter particles have $\beta = -0.2$ while those of the
host galaxy's halo have $\beta = 0.2$.  Such a scenario is consistent with the charge separation described in \cite{STcosmo} whereby dark-matter particles
of one sign are pushed away from the cores of initially neutral galaxies and subsequently collapse into satellites.  In this case the dark-matter forces
would be attractive {\it within} each individual halo but repulsive {\it between} the halos of the satellite and host galaxies.  We see in
Fig.~\ref{F:neg} that this leads to an asymmetry in the tidal streams opposite to what we have previously seen for purely attractive forces.  The left panel
shows that stellar densities have been enhanced in the leading stream and reduced in the trailing stream, in contrast to the center panel with no
dark-matter forces and the right panel where attractive forces lead to the opposite result.

\section{Discussion} \label{S:conc}

\begin{figure}
\scalebox{0.90}{\includegraphics{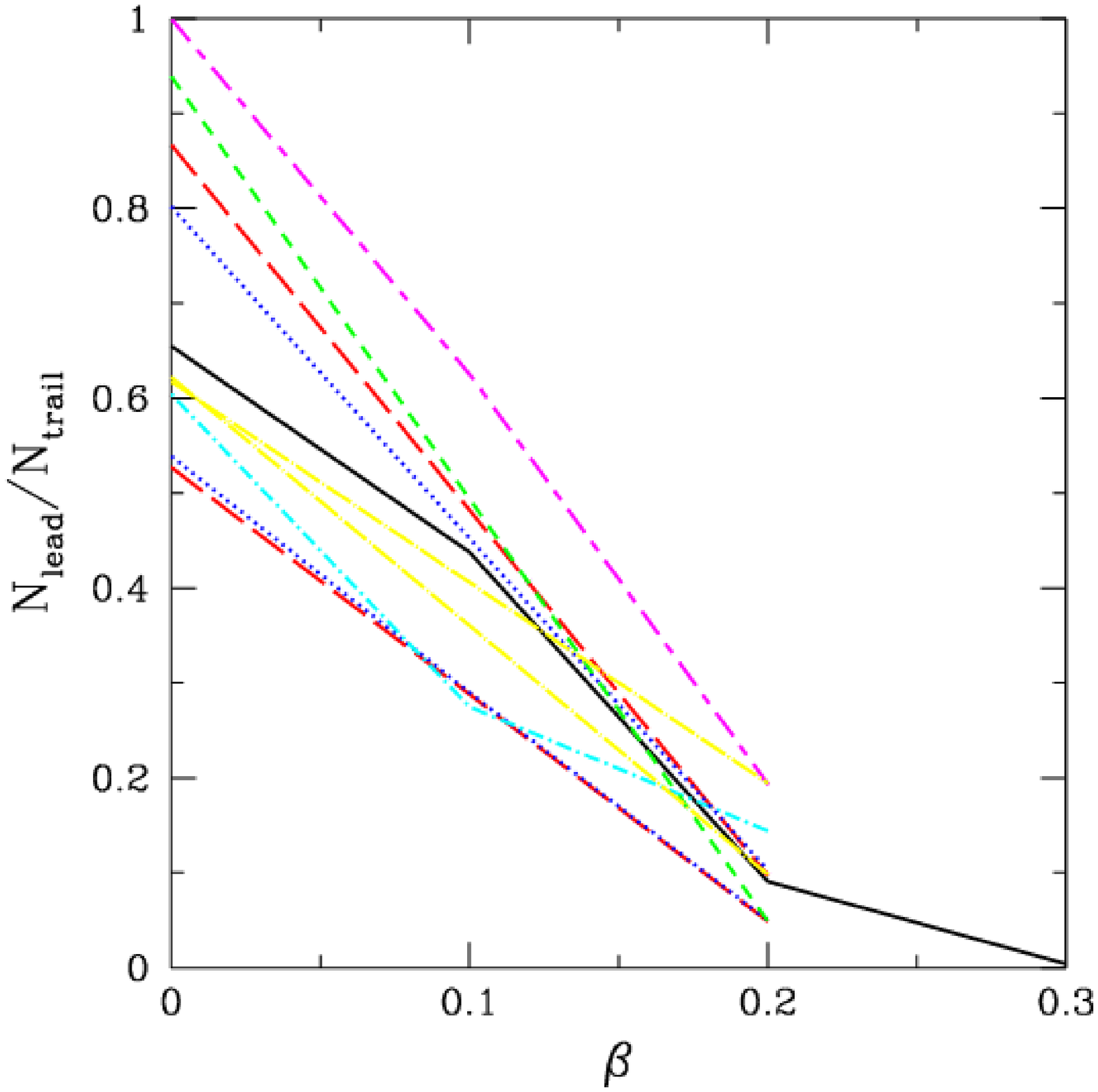}}
\caption{The ratio of leading to trailing stars as a function of charge-to-mass ratio $\beta$ for different models of the host-satellite system.  The black
(solid) curve is our default best-fit model (runs 1a-1d). The magenta (long-short dashed) curve doubles the initial mass of the satellite (runs 2a-2c).   The
two red (long-dashed) curves have satellites with spin angular momentum: the top curve has prograde spin (runs 3a-3b), while the bottom curve has retrograde
spin (runs 3c-3d).  The blue (dotted) curves have satellites with different orbits: the top curve has the satellite on a more circular orbit (runs 4a-4b),
while the bottom curve has the satellite on a planar orbit rather than the polar orbit of Sgr (runs 5a-5c).  The cyan (dot-short dashed) curve uses a Milky
Way model with lighter halo and heavier disk (runs 6a-6c).  The green (short-dashed) curve has a satellite where 25\% of the most bound particles represent
stars (runs 7c-7d).  The yellow (dot-long dashed) curves have satellites with lower $M/L$ ratios (higher stellar mass fractions): the top curve has
$M/L = 4.5$ while the bottom curve has $M/L = 10$.} \label{F:rat}
\end{figure}

In a recent {\it Letter} \cite{PRL}, we introduced the idea that an asymmetry in the leading and trailing streams of a tidally disrupting satellite could
be a signature of a dark-matter-only force.  Such a force cannot be detected in a system in isolation, as the stars we observe are only sensitive to the
dark-matter density profile which can be maintained in the presence of an attractive (repulsive) dark-matter force by increasing (decreasing) the velocities
of dark-matter particles.  However for satellites orbiting a larger host system, a dark-matter force will displace the center of mass of the satellite's
dark matter with respect to that of its stars leading to several potentially observable effects.  Previous work suggested that this displacement will produce
asymmetric rotation curves or warping in a satellite galactic disk \cite{EPlong}.  Unfortunately, not all satellites have well-defined disks and we would
certainly expect to observe warping in a strong tidal gravitational field.  Tidal streams provide a uniquely sensitive test of a dark-matter force because
in a system {\it already} disrupting due to gravitational effects, the dark-matter force tips the balance between the tidal bulges on the near and far side
of the satellite.  Our primary concern in this paper is to make sure that some other more conventional change to our models does not produce the same effect.

To accomplish this, we explored the specifics of tidal disruption in much greater detail and examined what factors
besides a dark-matter force influence the extent of tidal streams.  Features of galaxy morphology such as tidal streams have traditionally been viewed as
the messy products of nonlinear dynamics, a far cry from clean probes of cosmology like the cosmic microwave background (CMB).  While it may be difficult to
derive the detailed structure of an individual stream from first principles, tidal streams can be understood in general using analytic arguments supported
by numerical simulations.  With the discovery of several new stellar tidal streams possibly associated with Milky Way dwarf satellites
\cite{Willman,G&O,Grill}, and the prospect of further discoveries by SDSS and future survey missions like Gaia or SIM, we believe that serious consideration
of tidal streams as probes of long-range dark-matter interactions is warranted.

The wide variety of dark-matter candidates and possible dark-matter interactions makes it difficult to establish precise constraints, but any effective
violation in the universality of free fall between baryons and dark-matter on scales of $\mathcal{O}$(1 kpc) should conceivably be reflected in the
morphology of tidal streams.  Many other factors besides a dark-matter force affect the amount of tidal disruption, and in Section~\ref{S:res} we sought to
survey some the most important of these factors.  The results of this survey are summarized in Fig.~\ref{F:rat}.  We have counted the number of leading
stellar particles $N_{\rm lead}$ and trailing stellar particles $N_{\rm trail}$ and plotted their ratio as a function of $\beta$ for the simulations listed
in Table~\ref{T:runs}.  In most of the simulations, we have counted the stellar particles in $100^{\circ}$ slices of angular separation $\theta$ from the
satellite core as shown on the $x$-axis of Fig.~\ref{F:SD1}.  The position of these slices was chosen to include the stars piled up at the apocenters
furthest along the leading and trailing streams.  There were no prominent accumulations near apocenter in the simulations with a nearly circular orbit (runs
4a-4b) and those in which the stellar particles were 25\% of the the most bound particles (runs 7c-7d), so we counted all stellar particles further than
$10^{\circ}$ from the satellite core.  Runs 7a and 7b were not presented on this plot as there was virtually no tidal disruption in these simulations.

At first glance, Fig.~\ref{F:rat} confirms our worst suspicions of tidal streams as a probe of cosmology.  In the absence of a dark-matter force
$(\beta = 0.0)$, the leading-to-trailing ratio still varies by almost a factor of two between simulations.  Changes like doubling the satellite mass or
adding prograde spin dramatically increase the total amount of tidal disruption and increase the leading-to-trailing ratio to some extent as well.  In
contrast, retrograde spin and a planar orbit reduce the leading-to-trailing ratio compared to the default simulations shown by the black solid curve.
However, in spite of all these changes, the leading-to-trailing ratio exceeds 0.5 for {\it all} simulations without a dark-matter force and {\it never}
exceed 0.2 for $\beta = 0.2$, a dark-matter force only 4\% the strength of gravity.  While it would be premature to conclude that a dark-matter force exists
based on a single number, measurements of the satellite core and stellar distances and radial velocities along the streams should tell us about what
leading-to-trailing ratio to expect.  An observed leading-to-trailing ratio significantly below (above) this value would alert us to the possible presence
of an attractive (repulsive) dark-matter force.  Even if observations of a single tidal stream did not convince us of the existence of a dark-matter force,
several tidal streams have already been discovered about the Milky Way.  A consistent asymmetry observed in different satellites on different orbits might
be quite persuasive.

Before concluding, we wish to make a brief comment in the implications of modified Newtonian dynamics (MOND) for tidal streams.  The model explored in this
paper is not an example of MOND; it includes copious amounts of dark matter and explains all conventional dark-matter signatures (e.g. galactic rotation
curves, cluster velocities dispersions) through the Newtonian gravitational influence of this dark matter.  Nonetheless, one might imagine that the Sgr
tidal streams could place interesting constraints on MOND, and indeed this was the subject of a recent paper \cite{ReadMOND}.  That work focused on the
precession of the Sgr orbital plane; naively one would expect MOND to overpredict the amount of precession as the potential would be less spherically
symmetric in the absence of a dark-matter halo.  However, the potential sourced by an infinitely thin disk is much more spherical in MOND than Newtonian
gravity, implying an amount of precession similar to that of a mildly oblate dark-matter halo $(q = 0.9)$. MOND could not explain the discrepancy between Sgr
leading-stream velocity data, which favors a prolate halo, and the orbital precession which prefers an oblate halo \cite{2MassTT}.  We considered MOND using
the analytical argument presented in Section~\ref{S:tidal}, with an acceleration proportional to $1/r$ below the MOND scale
$a_0 = 1.2 \times 10^{-10}$ m/s$^2$.  This flatter potential led to a slightly larger natural asymmetry $\Delta r_{\rm nat}$ despite the smaller ratio
$m_{\rm sat}/M_R$ in the absence of dark matter.  MOND does not provide a mechanism for the strong suppression of the leading (trailing) tidal stream we
expect for an attractive (repulsive) dark-matter force for $\beta \gtrsim 0.2$.  Our model also does not seem to offer an explanation for the
halo-oblateness discrepancy for acceptable values of $\beta$, though we have not explored what a careful conspiracy of parameters might allow.

Although we believe using tidal streams to constrain a dark-matter force is a promising approach, significant observational and theoretical challenges
remain before it can be practically implemented.  Accurate measurements of the satellite core are essential to determining its mass, mass-to-light ratio,
and possible spin.  We must also find a reliable method of identifying stream stars at large separations $\theta$ from the satellite core, and must take
care that this method does not introduce a bias between trailing and leading stars.  On the theoretical front, more work needs to be done to establish a
correspondence between data and simulations.  It may not be feasible to evolve streams backwards into a bound satellite, so we need to develop techniques to
guess the appropriate initial conditions more systematic than trial and error.  We hope to confront these issues in the course of searching for a
dark-matter force in actual observations of the Sgr tidal stream.  While neither theory nor observation offer as yet a compelling reason to believe in the
existence of a dark-matter force, the tremendous payoff the discovery of such a force could provide in terms of learning physics beyond the Standard Model
makes the search well worth the effort.

\begin{acknowledgments}
We wish to thank Professors John Dubinski and Larry Widrow for assistance with the use of their code GALACTICS, and CITAzens Pat McDonanld, Neal Dalal,
Christoph Pfrommer, and Jonathan Sievers for useful conversations.
All computations were performed on CITA's McKenzie cluster \cite{McKenzie}, which was funded by the Canada Foundation for Innovation and the Ontario
Innovation Trust.  Kesden acknowledges support from the NASA Graduate Research Program, and the National Sciences and Engineering Research Coucil (NSERC) of
Canada.  Kamionkowski acknowledges support from the DoE DE-FG03-92-ER40701, NASA NNG05GF69G, and the Gordon and Betty Moore Foundation.
\end{acknowledgments}

\bibliography{Tidal830}% Produces the bibliography via BibTeX.

\end{document}